\documentclass[conference,twocolumn]{sig-alternate-10pt}
\usepackage{verbatim,epsfig,times,subfigure}
\usepackage{amsmath,amsfonts,amssymb,mathrsfs}
\usepackage{url}

\newtheorem{definition}{{\bf \hspace{-0.18in} Definition}}
\newtheorem{assumption}{{\bf \hspace{-0.18in} Assumption}}
\newtheorem{theorem}{{\bf \hspace{-0.18in} Theorem}}
\newtheorem{corollary}{{\bf \hspace{-0.18in} Corollary}}
\newtheorem{lemma}{{\bf \hspace{-0.18in} Lemma}}
\def\done{\hspace*{\fill} \rule{1.8mm}{2.5mm}}

\begin{document}
\title{Subsidization Competition: Vitalizing the Neutral Internet}

\numberofauthors{1}
\author{
\alignauthor
Richard T. B. Ma\\
              \affaddr{National University of Singapore and}
                     \affaddr{Advanced Digital Science Center}\\
       \email{tbma@comp.nus.edu.sg}
}

\maketitle


\begin{abstract}
Unlike telephone operators, which pay {\em termination fees} to reach the users of another network, Internet Content Providers (CPs) do not pay the Internet Service Providers (ISPs) of users they reach.
While the consequent cross subsidization to CPs has nurtured content innovations at the edge of the Internet, it reduces the investment incentives for the access ISPs to expand capacity.
As potential charges for terminating CPs' traffic are criticized under the net neutrality debate, we propose to allow CPs to voluntarily subsidize the usage-based fees induced by their content traffic for end-users.
We model the regulated subsidization competition among CPs under a neutral network and show how deregulation of subsidization could increase an access ISP's utilization and revenue, strengthening its investment incentives.
Although the competition might harm certain CPs, we find that the main cause comes from high access prices rather than the existence of subsidization.
Our results suggest that subsidization competition will increase the competitiveness and welfare of the Internet content market; however, regulators might need to regulate access prices if the access ISP market is not competitive enough.
We envision that subsidization competition could become a viable model for the future Internet.
\end{abstract}


\section{Introduction}
Driven by bandwidth-intensive video applications and pervasive Internet access via mobile devices, Internet inter-domain traffic has been growing more than $50\%$ per annum \cite{craig10internet} and is expected to grow more than $100\%$ per annum for mobile networks \cite{ATKEARNEY}.
To sustain such rapid traffic growth, Internet Service Providers (ISPs), especially the mobile access ISPs, need to upgrade their network infrastructures and expand capacities.
However, revenues from online services are growing more than twice as fast as those from Internet access \cite{ATKEARNEY}, whose market capitalizations have stagnated as investors weigh high capital requirements against continued margin pressure. This raises serious challenges regarding the viability of the current Internet model in the future.
\begin{figure}[h]
\centering
\includegraphics[width=.99\columnwidth]{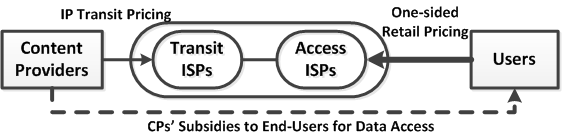}
\caption{Subsidization in a two-sided Internet market.}
\label{figure:two-sided_model}
\end{figure}
The fundamental problem lies in the two-sided market structure \cite{Armstrong06,rochet03platform} of the Internet, where the ISPs provide a platform connecting end-users to the Internet Content Providers (CPs).
The CP-side transit ISPs are often the backbone ISPs that provide IP transit services. Due to market competition, recent years have witnessed a constant decrease in IP transit prices \cite{Norton}.
The more serious problem lies in the last-mile connections toward the end-users, which are often the bottlenecks of the Internet \cite{weber-pricing}. For these last-mile access ISPs, including the mobile providers, the main source of revenue is from their end-users. Traditionally, they also need to pay transit ISPs for transit services, although settlement-free peering is common in practice. 
Existing proposals to resolve this issue include: 1) allowing ISPs to impose termination fees \cite{apple10Bloomberg} on CPs' traffic towards their end-users, and 2) allowing ISPs to prioritize traffic, differentiate services and thus charge users non-uniformly.
However, both proposals have triggered a heated debate over network neutrality \cite{tim05nn,jon07net}, whose advocates argue that {\em zero-pricing} \cite{lee09subsidizing} and the neutrality of physical networks are needed to protect and encourage content innovations of the Internet. 
After all, the limited capacities and rising access costs will dampen user demand, if CPs cannot return the created value to ISPs.

Traditionally, wireline access ISPs charge flat-rate prices and implicitly subsidize CPs for content delivery; however, usage-based pricing is commonly used by wireless providers. The FCC chairman has recently backed usage-based pricing for broadband to penalize heavy Internet users \cite{Johnson13fcc,schatz10fcc} and major U.S. broadband providers, e.g., Verizon \cite{Segall} and AT\&T \cite{Taylor}, have adopted tiered schemes where charges are imposed to the metered usage above a predefined data cap.
We assume the adoption of usage-based pricing by access ISPs, under which heavy users share the burden of subsidizing CPs.
Our goal is to realign the created value and stimulate the needed investment in the Internet value chain and our high-level idea is illustrated in Figure \ref{figure:two-sided_model}, where we propose to allow CPs to voluntarily subsidize the usage-based access costs caused by their content traffic, partially or fully, for the end-users.
Our solution modularizes the issue along the tussle boundary such that the physical network is kept neutral while innovative competition among CPs is enabled at higher service layers so as to incentivize ISP investments.
As a special case of full subsidization, AT\&T's recent {\em sponsored data} \cite{ATT14} plan will likely be available by the end of 2014. Nevertheless, it raised concerns over anti-competitive practices, and the FCC says that they will be monitoring and prepared to intervene if necessary \cite{Nagesh14}.

To understand and demonstrate the potential impact of subsidization, we take an analytical approach as no market data is available for an empirical study yet.
We develop a macroscopic Internet model and study the status quo one-sided pricing of the access ISPs in Section \ref{sec:model}. Building upon this framework, we model the policy that regulates subsidization and analyze the competition game among CPs in Section \ref{sec:competition}. We evaluate the policy implications on the ISP's revenue and system welfare in Section \ref{sec:revenue} and discuss the regulatory and implementation issues, as well as the limitations of our model in Section \ref{sec:issues}.
Our analytical results include:
\begin{itemize}
\item We characterize the impact of capacity, user population (Theorem \ref{theorem:capacity_user_base_effect}) and one-sided ISP pricing (Theorem \ref{theorem:price_effect}) on the system utilization and the CPs' throughput.

\item We analyze the CPs' subsidization competition game and characterize the Nash equilibrium (Theorem \ref{theorem:Nash_equilibrium}), its uniqueness and its dynamics (Theorem \ref{theorem:unique_Nash_equilibrium} and \ref{theorem:equilibrium_dynamics}).

\item We characterize the impact of subsidization policy on the CPs' throughput,  user population, and the system's utilization (Theorem \ref{theorem:regulation_effect}) and welfare (Corollary \ref{corollary:welfare}).
\end{itemize}

Through our theoretical results and qualitative evaluations, we also identify important policy implications as follows.
\begin{itemize}
\item If allowed, profitable CPs tend to subsidize users more (Theorem \ref{theorem:subsidy_valuation_monotonicity}) and achieve higher throughput (Lemma \ref{lemma:subsidy_effect}), leading to higher system welfare (Corollary \ref{corollary:welfare}).

\item For fixed ISP pricing, deregulation of subsidization will increase the utilization and revenue of the access ISPs (Corollary \ref{corollary:utilization_monotonicity}), strengthening their investment incentives.

\item Although the throughput of certain CPs might decrease, this mainly comes from the ISP's high price (Theorem \ref{theorem:regulation_effect}) rather than the deregulation of subsidization.
\end{itemize}

In short, subsidization provides a feedback channel for the highly profitable CPs to repay their users and indirectly transfer value to the access ISPs.
It helps the access ISPs attract investment for capacity expansion while keeping their physical networks neutral.
Our results suggest that regulators should promote this subsidization competition; however, they might need to regulate the ISP's pricing if the access market is not competitive enough. 
We envision that subsidization competition could become a viable model for the future Internet so as to accommodate the growing demand for data traffic.


\section{Related Work}
A report from A.T. Kearney \cite{ATKEARNEY} studied the performance, economic and policy pressures of the Internet and identified four models for the future Internet: 1) modification of retail pricing schemes, 2) traffic-dependent charges, 3) enhanced-quality services over the public Internet, and 4) bilateral service agreements.
We discuss the above existing models and their relationships with our subsidization competition model.

\subsection{Modification of Retail Pricing Schemes}
Historically, the Internet adopted flat-rate prices \cite{odlyzko01internet,shakkottai08simplicity} for simplicity.
Economists \cite{mackiemason95} and computer scientists \cite{courcoubetis00, honig95usage, kesidis08flat} advocated usage-based pricing, which was shown to provide congestion control \cite{honig95usage}, quality of service \cite{kesidis08flat} and economic efficiency \cite{courcoubetis00}, and is adopted by mobile providers \cite{morgan11}.
Subsidization competition assumes the adoption of usage-based pricing by access ISPs, under which the consumers' usage-based charges could be subsidized by the CPs.


\subsection{Imposition of Termination Fees on CPs}
One core question under the net neutrality debate \cite{tim05nn} is whether ISPs should be allowed to use two-sided pricing and charge CPs for terminating their content traffic.
Prior work \cite{choi10net, musacchio09network, Njoroge10investment} studied the investment incentives in Internet-like two-sided markets and drew different conclusions.
Njoroge et al. \cite{Njoroge10investment} found that through CP-side pricing, ISPs could extract higher surplus and maintain higher investment levels. However, Choi et al. \cite{choi10net} found that expending capacity will decrease the CP-side sale price. Musacchio et al. \cite{musacchio09network} found that although the returns on investment under one-sided or two-sided pricings are comparable, the size of investment and profit depend on the advertising rates and users' price sensitivity.
From an economic efficiency point of view, \cite{Njoroge10investment} concluded that two-sided pricing results in higher social welfare; however, Lee and Wu \cite{lee09subsidizing} argued that the {\em zero pricing} at the CP-side
could be optimal in theory \cite{Armstrong06,rochet03platform}. Also, \cite{choi10net} found that the short-run welfare is higher under one-sided neutral regulation.
Our subsidization mechanism does not impose charges on the CPs by the ISPs, but provides an indirect mechanism for them to voluntarily return the {\em zero-pricing subsidy} back to the ISPs via their end-users.

\subsection{Differentiated Services over the Internet}
Another core debate under net neutrality is whether ISPs should be allowed to differentiate services and prices. Wu \cite{tim05nn} surveyed the discriminatory practices of broadband operators and proposed solutions to manage bandwidth and police ISPs. Sidak \cite{sidak06consumer} focused on consumer utility and argued that differential pricing is essential to maximize utility. Shetty et al. \cite{shetty10internet} considered a two-class service model and studied capacity planning, regulation, as well as differentiated pricing to consumers.
Economides et al. \cite{economides12network} compared various regulations for quality of service and price discrimination, and drew conclusions about desirable regulation regimes.
As net neutrality is undergoing heated debate, regulation of desirable differentiation is a rich area for further study. Under our solution, ISPs do not make any physical or economic differentiation and CPs' subsidies are made voluntarily. Subsidization serves as a policy mechanism that encourages competition and provides investment incentives and economic efficiency for the Internet ecosystem.

\subsection{Bilateral Settlements for Valued Services}
A fundamental cause of investment shortage is the misalignment between the value created and the revenue obtained in the Internet value chain. Clark et al. \cite{clark05tussle} discussed various economic tussles among the ISPs, CPs and users. 
Ma et al. \cite{ma10internet} advocated the use of the Shapley profit-sharing mechanism for multi-lateral ISP settlements and studied bilateral implementations \cite{ma11on}. 
Laskowski \cite{Laskowski06} identified that the network's lack of accountability is an obstacle to implement desirable contracting systems for the Internet.
Because accounting of traffic towards the end-users is feasible for access ISPs, our subsidization mechanism can be regarded as a contracting system between the CPs and access ISPs, even through they might not be directly connected to each other.
Through the feedback channel from CPs to end-users, subsidization provide a new avenue for the Internet ecosystem to realign the profit distribution among  different stakeholders.
\section{Macroscopic Internet Model}\label{sec:model}
To understand the implications of policy on the Internet, we start with a macroscopic model that captures the physical and business dynamics in the status quo Internet among an access ISP, its users and the CPs used by these consumers.

\subsection{Basic Physical System Model}
We denote $\mu$ as the capacity of an access ISP serving the users of a certain geographical region and $\cal N$ as the set of CPs used by the ISP's consumers.
For each CP $i\in\cal N$, we denote $m_i$ as its user population and define $\theta_i \triangleq m_i\lambda_i$ as its total throughput, where $\lambda_i$ denotes the CP's average per user throughput.
We define $\theta \triangleq \sum_{k\in\cal N}\theta_k$ as the aggregate system throughput.
Under any fixed capacity $\mu$ and aggregate throughput $\theta$, we define $\phi \triangleq \Phi(\theta, \mu)$ as the system utilization, a function of $\theta$ and $\mu$.
We assume the average user throughput is a function of the utilization, i.e., $\lambda_i \triangleq \lambda_i(\phi)$.

{\assumption $\Phi(\theta, \mu)$ and any $\lambda_i(\phi)$ are differentiable. $\Phi(\theta,\mu)$ is strictly decreasing in $\mu$, strictly increasing in $\theta$ and satisfies $\lim_{\theta\rightarrow 0} \Phi(\theta, \mu) = 0$ for all $\mu>0$. $\lambda_i(\phi)$ is strictly decreasing in $\phi$ and satisfies $\lim_{\phi\rightarrow \infty} \lambda_i(\phi)=0$. \label{assumption:utilization_throughput}}

Assumption \ref{assumption:utilization_throughput} captures the physics of capacity utilization: utilization increases when the system accommodates higher throughput or has less capacity, and vice-versa. Because the system utilization and congestion are two sides of the same coin (congestion happens when the system is over-utilized), 
it also captures the physics of user throughput: users achieve higher throughput when the system is less congested.

\begin{figure}[!h]
\centering
\includegraphics[width=.99\columnwidth]{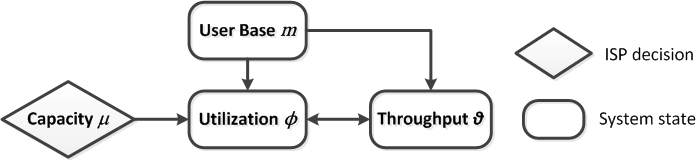}
\caption{The basic system model $({\bf m}, \mu)$.}
\label{figure:network_model}
\end{figure}

We denote $\mathbf{m}$ as the vector of user populations of the CPs. Figure \ref{figure:network_model} illustrates our basic model $({\bf m},\mu)$, where
the user populations ${\bf m}$ and system capacity $\mu$ collectively determine the system utilization $\phi$ and each CP $i$'s throughput $\theta_i$.
Because utilization increases with the accommodated throughput, which decreases with the utilization (congestion), the resulting
system utilization should be in an equilibrium. 


{\definition[Utilization] $\phi$ is the utilization of a system $(\mathbf{m}, \mu)$ if it satisfies the following condition:
\begin{equation}\phi = \Phi\left(\sum_{k\in\cal N}m_k\lambda_k(\phi), \mu\right).\label{equation:utilization}\end{equation}
\label{definition:utilization}}

Definition \ref{definition:utilization} states that the utilization $\phi$ should induce the aggregate system throughput $\theta = \sum_{k\in\cal N}m_k\lambda_k(\phi)$ such that it leads to exactly the same level of utilization $\phi = \Phi(\theta,\mu)$.

We define $\Theta(\phi, \mu) \triangleq \Phi^{-1}(\phi, \mu)$ as the inverse function of $\Phi(\theta, \mu)$ with respect to $\phi$. $\Theta(\phi, \mu)$ can be interpreted as the implied amount of throughput that induces a utilization level $\phi$ for a system with capacity $\mu$. By Assumption \ref{assumption:utilization_throughput}, $\Theta(\phi, \mu)$ is strictly increasing in both $\phi$ and $\mu$. To characterize the utilization of a system $(\mathbf{m}, \mu)$, we define a {\em gap} function $g(\phi)$ between the supply and demand of throughput under a fixed level of utilization $\phi$ as
$g(\phi) \triangleq \Theta\left(\phi,\mu\right) - \sum_{k\in\cal N}m_k\lambda_k(\phi)$.

{\lemma Given fixed capacity $\mu$ and user populations $\mathbf{m}$, $g(\phi)$ is a strictly increasing function of $\phi$.
The system operates at a unique level of utilization $\phi$, which solves $g(\phi)=0$.\label{lemma:unique_utilization}}

Lemma \ref{lemma:unique_utilization} characterizes the uniqueness of the system utilization $\phi$ under which the throughput supply $\Theta(\phi,\mu)$  equals the aggregate throughput demand $\sum_{k\in\cal N}m_k\lambda_k(\phi)$.
Based on Lemma \ref{lemma:unique_utilization}, for any system $(\mathbf{m}, \mu)$, we denote $\phi(\mathbf{m}, \mu)$ as its unique system utilization and $\theta_i(\mathbf{m}, \mu)$ as CP $i$'s corresponding throughput, i.e., $\theta_i(\mathbf{m}, \mu) \triangleq m_i\lambda_i\left(\phi(\mathbf{m}, \mu)\right) $.

We define the marginal change of the {\em throughput gap} $g(\phi)$ due to the marginal change in the system utilization $\phi$ as
\begin{equation}
\frac{dg}{d\phi}\triangleq\frac{\partial \Theta(\phi,\mu)}{\partial \phi}-\sum_{k\in\cal N}m_k\frac{d\lambda_k(\phi)}{d\phi} > 0,\label{equation:dg}
\end{equation}
where the first term (second summation) captures the change of throughput in supply (total demand).
We characterize the impact of user population $m_i$ and capacity $\mu$ on the utilization $\phi$ and throughput $\theta_i$, as a function of $dg/d\phi$ as follows.

{\theorem[Capacity and User Effect] The system utilization $\phi$ decreases if the capacity $\mu$ increases or any user population $m_i$ decreases. In particular, we have
\begin{equation} \frac{\partial \phi}{\partial \mu} = - \left( \frac{dg}{d\phi} \right)^{-1} \frac{\partial \Theta(\phi,\mu)}{\partial \mu}<0, \label{equation:phi_mu}
\end{equation}
\begin{equation} \text{and} \quad \frac{\partial \phi}{\partial m_i} =  \left(\frac{dg}{d\phi}\right)^{-1}\lambda_i>0, \quad \forall i\in\cal N.\label{equation:phi_mi}
\end{equation}
Any CP $i$'s throughput $\theta_i$ increases with capacity $\mu$ and its user population $m_i$; however, the increase in $m_i$ reduces the throughput $\theta_j$ of all other CPs $j\neq i$. In particular,
\[ \frac{\partial \theta_i}{\partial \mu}=m_i\frac{d\lambda_i}{d\phi}\frac{\partial \phi}{\partial m_i} >0, \quad \frac{\partial \theta_i}{\partial m_i} = \lambda_i+m_i \frac{d\lambda_i}{d\phi}\frac{\partial \phi}{\partial m_i}>0 \]
\[  \text{ and} \quad \frac{\partial \theta_j}{\partial m_i}=m_j \frac{d\lambda_j}{d\phi}\frac{\partial \phi}{\partial m_i}<0, \quad \forall j\neq i.\]
\label{theorem:capacity_user_base_effect}}

Theorem \ref{theorem:capacity_user_base_effect} intuitively states that 1) when the system capacity becomes more abundant, the throughput of any CP increases and the system utilization decreases, and 2) when more users of a CP join the system, the system utilization and the throughput of that CP increase, while that of any other CP decreases.
Equation (\ref{equation:phi_mu}) explicitly shows that the capacity effect on the system utilization can be expressed as its impact on the feasible system throughput $\partial \Theta(\phi,\mu)/\partial \mu$, normalized by $dg/d\phi$, which re-balances the supply and demand of throughput so as to fill the {\em throughput gap} $g(\cdot)$ under the new system utilization. Similarly in Equation (\ref{equation:phi_mi}), $\lambda_i$ can be interpreted as the marginal increase in throughput when a new user of CP $i$ joins the system. It also implies that the user impact on the utilization is proportional to the per user throughput, i.e., ${\partial \phi}/{\partial m_i}:{\partial \phi}/{\partial m_j}=\lambda_i:\lambda_j$.


CPs' throughput depends on whether their users are sensitive to congestion. This can be characterized by {\em utilization-elasticity of throughput} or {\em $\phi$-elasticity of $\lambda_i$}, defined as $\epsilon_{\phi}^{\lambda_i}$.


{\definition[Elasticity] The elasticity of $y$ with respect to $x$, or $x$-elasticity of $y$, $\epsilon_x^y$ is defined as
$\displaystyle\epsilon_x^y \triangleq \frac{\partial y}{\partial x}\frac{x}{y}$.\label{definition:elasticity}}

Elasticity is often expressed as $\epsilon_x^y = ({\partial y}/y)/({\partial x}/x)$ and interpreted as
the percentage change in $y$ (the numerator) in respond to the percentage change in $x$ (the denominator).
In particular, $\epsilon_{\phi}^{\lambda_i}$ captures the percentage change in $\lambda_i$ in respond to the percentage change in the system utilization $\phi$.

{\lemma If CP $i$ is replaced by a new CP $j$ which satisfies $m_j \lambda_j(0) = m_i \lambda_i(0)$ and $\epsilon_{\phi}^{{\lambda}_j}(\phi) = \epsilon_{\phi}^{{\lambda}_i}(\phi)$ for all $\phi>0$, then the system utilization does not change. \label{lemma:CP_characteristics}}

Lemma \ref{lemma:CP_characteristics} characterizes CPs' throughput by $\phi$-elasticity.
Mathematically, by keeping the same elasticity $\epsilon_{\phi}^{{\lambda}_i}$, we could scale up the maximum throughput $\lambda_i(0)$ by $\tilde\lambda_i(0) =\kappa \lambda_i(0)$ and scale down $m_i$ by $\tilde m_i = \kappa^{-1}m_i$ for some constant $\kappa>0$ without affecting the system utilization $\phi$ and the throughput of all other CPs. Operationally, it implies that we could treat the traffic of a CP $i$ as if it has a single big user, i.e., $\tilde m_i=1$, with the maximum throughput $\tilde\lambda_i(0) = m_i \lambda_i(0)$. It also implies that for multiple CPs, e.g., a set $\cal S \subset \cal N$, with the same $\phi$-elasticity of throughput, we could conceptually aggregate them as a single CP $k$ with parameters $m_k$ and $\lambda_k(0)$ that satisfy $m_k \lambda_k(0) = \sum_{i\in\cal S} m_i \lambda_i(0)$.
By this result, we will use a single CP to model a group of CPs that have similar traffic characteristics in our later numerical evaluations.

\subsection{One-Sided ISP Pricing}
In the current Internet, most access ISPs only charge their end-users. We denote $t_i$ as the per-unit usage charge for the data traffic of CP $i$. We assume that CP $i$'s user population $m_i$ is a function of its per-unit charge $t_i$, i.e., $m_i=m_i(t_i)$.
{\assumption $m_i(t_i)$ is continuously differentiable and decreasing with $\lim_{t_i\rightarrow \infty} m_i(t_i)=0$. \label{assumption:user_base}}

Assumption \ref{assumption:user_base} states that the user demand decreases with the per-unit charge on the content traffic. This model is quite general and coincides with existing models \cite{Armstrong06,rochet03platform} where the heterogeneity of users are modeled by the distribution of their valuations on data traffic and only the users whose valuations are higher than $t_i$ consume data traffic from CP $i$.

Because traffic from different CPs should not be differentiated under net neutrality, ISPs often impose a uniform charge. We denote $p$ as this uniform charge for all types of traffic. When CPs do not provide subsidies, we have $t_i=p$ for any $i\in \cal N$.
We define the ISP's revenue as $R \triangleq p\theta = p\sum_{i\in\cal N}m_i\lambda_i$ and each CP $i$'s utility as $U_i = v_i\theta_i$, where $v_i$ denotes CP $i$'s average per-unit traffic profit.
Because the price $p$ affects the user demand $\mathbf{m}$, we extend our basic system model to capture the one-sided ISP pricing as follows.
\begin{figure}[!h]
\centering
\includegraphics[width=.99\columnwidth]{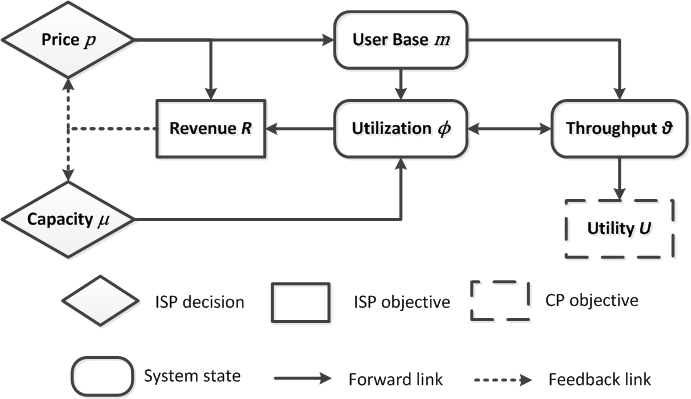}
\caption{The system model with one-sided ISP pricing.}
\label{figure:one_sided_model}
\end{figure}

Figure \ref{figure:one_sided_model} shows that the user populations $\mathbf{m}$ are influenced by the price $p$, and that the utilization $\phi$ affects the ISP's revenue $R = p\sum_{i\in\cal N}m_i\lambda_i(\phi)$. 
ISPs can use revenue as feedback to guide their pricing and capacity decisions; however,  CPs' utilities are determined by throughput, which they have no means to influence in the one-sided pricing model.

Because $t_i=p$ for any CP $i$, by Assumption \ref{assumption:user_base}, we express the user populations as a function of price as $\mathbf{m} = \mathbf{m}(p)$. For a fixed capacity $\mu$, we define $\phi(p) \triangleq \phi({\mathbf m}(p),\mu)$ as the utilization of system $(\mathbf{m}(p),\mu)$ and $\theta_i(p) \triangleq m_i(p)\lambda_i\left(\phi(p)\right)$ as the corresponding throughput of CP $i$ under price $p$. By Assumption \ref{assumption:utilization_throughput} and \ref{assumption:user_base}, both $\phi(p)$ and $\theta_i(p)$ are differentiable functions of $p$. We characterize the impact of price $p$ on the system utilization $\phi$ and each CP $i$'s throughput $\theta_i$ in the following theorem. The price's impact on $U_i$ is simply its impact on $\theta_i$ scaled by $v_i$ and we will study its impact on the revenue $R$ under a more general setting in Section \ref{sec:revenue}.

{\theorem[Price Effect] The system utilization $\phi$ decreases with the price $p$. In particular, we have
\begin{equation} \frac{\partial \phi}{\partial p} =  \left(\frac{dg}{d\phi}\right)^{-1}  \sum_{k\in\cal N} \frac{dm_k}{dp} \lambda_k \leq 0.\label{equation:phi_p}\end{equation}
The aggregate throughput $\theta$ decreases with the price $p$,
\begin{equation} \frac{\partial \theta}{\partial p} =  \frac{\partial \Theta}{\partial \phi}   \left(\frac{\partial \Theta}{\partial \phi} - \sum_{i\in\cal N}m_i \frac{d\lambda_i}{d\phi}\right)^{-1}  \sum_{i\in\cal N} \frac{dm_i}{dp} \lambda_i \leq 0, \label{equation:dtheta_dp}\end{equation}
and any CP $i$'s throughput $\theta_i$ increases with $p$ if and only if
\begin{equation}
 {\epsilon_p^{m_i}}/{\epsilon_\phi^{\lambda_i}} < - \epsilon_p^\phi.\label{inequality:price_effect}
\end{equation}
\label{theorem:price_effect}}
Theorem \ref{theorem:price_effect} states that when the price increases, due to its direct impact on decreasing the user demand, the system utilization decreases, which implies that the aggregate throughput of all the CPs also decreases. Similar to (\ref{equation:phi_mu}) and (\ref{equation:phi_mi}), the price effect on the system utilization can be expressed as its impact on the aggregate throughput $d\theta/dp$, normalized by $dg/d\phi$ in Equation (\ref{equation:phi_p}).
However, a CP $i$'s throughput might increase or decrease, depending on its utilization elasticity of (per user) throughput $\epsilon_\phi^{\lambda_i}$ and price elasticity of user demand  $\epsilon_p^{m_i}$, i.e., the percentage change in the population $m_i$ in response to  the percentage change in price $p$.
In particular, $\theta_i$ increases with $p$ if its users are less price-sensitive and more congestion-sensitive, i.e., $|{\epsilon_p^{m_i}}|$ is small and $|{\epsilon_\phi^{\lambda_i}}|$ is large. Notice that the right hand side of (\ref{inequality:price_effect}) measures the overall impact of price on the system utilization.

To understand the price effect more intuitively, we illustrate a numerical example as follows.
We consider a utilization function $\Phi(\theta,\mu) = \theta/\mu$, which uses {\em per capacity throughput} as the metric for system utilization.
We consider the forms of throughput $\lambda_i(\phi)=e^{-\beta_i \phi}$ and user demand $m_i(t_i) = e^{-\alpha_i t_i}$ that satisfy Assumption \ref{assumption:utilization_throughput} and \ref{assumption:user_base}.
Under these exponential forms, the $\phi$-elasticity of throughput $\epsilon_{\phi}^{\lambda_i}$ equals $-\beta_i \phi$ and the $p$-elasticity of user population $\epsilon_{p}^{m_i}$ equals $-\alpha_i p$.
Based on the above setting, we can express a CP $i$'s throughput as $\theta_i = e^{-(\alpha_i p+\beta_i\phi)}$, where $\alpha_i$ and $\beta_i$ neatly characterize $\theta_i$'s sensitivity to price and congestion, respectively.
We obtain ${dg}/{d\phi} = \mu + \sum_{i\in\cal N}\beta_i \theta_i$ and
\[ \frac{\partial \phi}{\partial p} = \left(\frac{dg}{d\phi}\right)^{-1} \left(\sum_{i\in\cal N} \frac{dm_i}{dp}\lambda_i\right) = -\frac{\sum_{i\in\cal N}\alpha_i \theta_i}{\mu+\sum_{i\in\cal N}\beta_i \theta_i}.\]
By Theorem \ref{theorem:price_effect}, 
throughput $\theta_i$ increases at price $p$ if 
\begin{equation} \frac{\alpha_i p}{\beta_i \phi} < \frac{\sum_{j\in\cal N}\alpha_j \theta_j}{\mu+\sum_{k\in\cal N}\beta_k \theta_k}. \label{equation:illustration}
\end{equation}
We set the system capacity to be $\mu=1$ and consider a set of $9$ types of CPs with values of $(\alpha_i,\beta_i)$ chosen from $\{1,3,5\}$.

\begin{figure}[!h]
\centering
\includegraphics[width=.99\columnwidth]{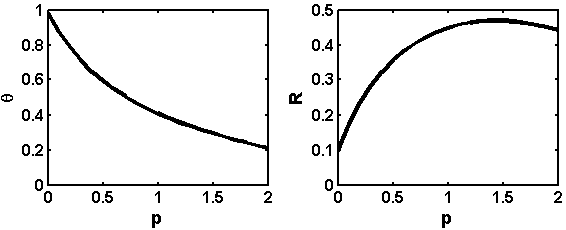}
\caption{Aggregate throughput $\theta$ and ISP's revenue $R$.}
\label{figure:R}
\end{figure}

Figure \ref{figure:R} plots the aggregate throughput $\theta$ (left) and the ISP's revenue $R$ (right) as a function of price $p$ that varies along the x-axis. We observe that the aggregate throughput decreases with the price as indicated by Theorem \ref{theorem:price_effect}; however, the revenue $R=p\theta$ depends on both the price and aggregate throughput and shows a single-peak pattern.

\begin{figure}[!h]
\centering
\includegraphics[width=.99\columnwidth]{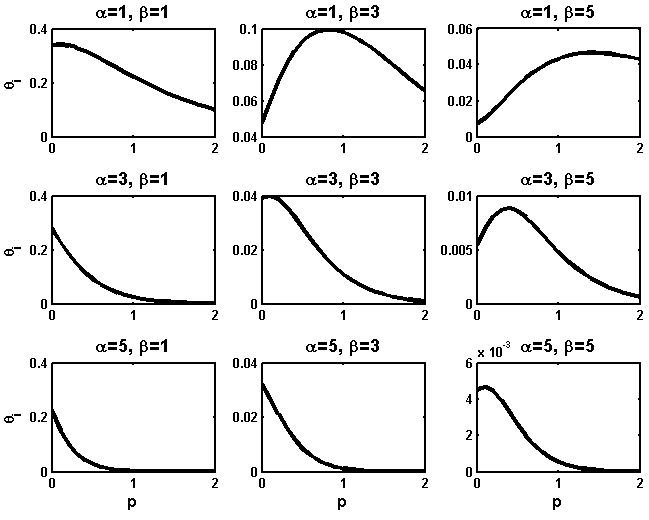}
\caption{Throughput $\theta_i$ of different CPs.}
\label{figure:theta_i}
\end{figure}

Figure \ref{figure:theta_i} shows the throughput $\theta_i$ of the $9$ individual CPs as a function of the price $p$ in each sub-figure, respectively.
In general, the throughput is low with large values of $\alpha_i$ and $\beta_i$ (the lower and right sub-figures), because the user population is more sensitive to price and the per user throughput is more sensitive to congestion.
By Theorem \ref{theorem:price_effect}, when $p$ increases, $\phi$ decreases, and therefore $(\alpha_i / \beta_i)(p/\phi)$ increases.
As indicated by condition (\ref{equation:illustration}), each $\theta_i$ decreases with $p$ eventually; however, when $p$ is small, we observe that the CPs with a small ratio of $\alpha_i/\beta_i$ (the upper and right sub-figures) demonstrate an increasing trend in throughput. Intuitively, for these CPs, the increase in the per user throughput $\lambda_i$ is much higher than the decrease in the user population $m_i$ so that the aggregate throughput $\theta_i$ could still increase.

{ \it \bf Regulatory Implications:}  Under the existing one-sided pricing, an increase in the access ISP's price reduces the user demand (by Assumption \ref{assumption:user_base}), and consequently reduces the system utilization and the total system throughput (by Theorem \ref{theorem:price_effect}). Regulators might want to regulate the price of an access ISP if its high price induces low utilization and drives the total system throughput too low (shown in Figure \ref{figure:R}).
However, as the existing wireless capacities of many carriers are often under-provisioned and highly-loaded, and regulations will further limit the ISPs' profit margin, we do not see the need for price regulation under the status quo.

\section{Subsidization Competition}\label{sec:competition}
As the access ISPs can neither differentiate services nor charge CPs in the one-sided pricing model in Figure \ref{figure:one_sided_model}, the limited profit margin does not provide enough investment incentive for them to expand capacities.
Neither do the CPs have any means to express preferences and improve their utilities.
In this section, we propose to create a feedback channel for the CPs to influence the system by allowing them to voluntarily subsidize the usage-based fees for their users.
\subsection{The Subsidization Competition Model}
We denote $q\geq 0$ as a subsidization policy that limits the maximum subsidy allowed.
We denote $s_i \in [0,q]$ as the per-unit usage subsidy provided by CP $i$ for its content traffic for users. We denote $\mathbf{s}$ as the vector of subsidies of the CPs. We extend the definition of CP $i$'s utility as $U_i \triangleq (v_i-s_i)\theta_i$.

\begin{figure}[!h]
\centering
\includegraphics[width=.99\columnwidth]{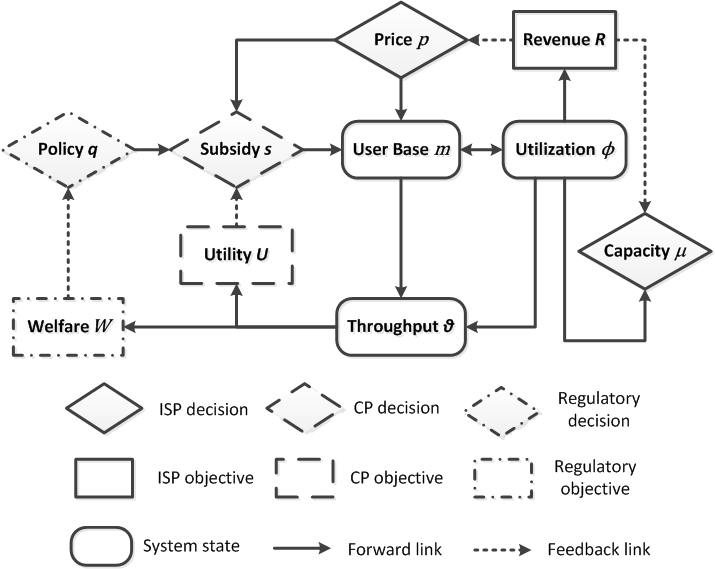}
\caption{One-sided ISP pricing with CP subsidization.}
\label{figure:network_model_with_subsidy_competition}
\end{figure}
Figure \ref{figure:network_model_with_subsidy_competition} illustrates the extended model where subsidization from CPs is allowed.
In this model, each CP $i$ could strategically choose its subsidy $s_i$ to influence its throughput $\theta_i$ via its user population $m_i$ so as to optimize its utility $U_i$.
Regulators can also use a certain welfare metric $W$ to determine the desirable policy regime $q$ for the Internet industry.

Under price $p$ and subsidies $\mathbf{s}$, user population $m_i$ satisfies $m_i(t_i) = m_i(p - s_i)$.
Given any fixed ISP decision $(p,\mu)$, we denote $\phi(\mathbf{s}) \triangleq \phi({\mathbf m}(p,\mathbf{s}),\mu)$ as the system utilization and define $\theta_i(\mathbf{s})=m_i(p-s_i)\lambda_i\left(\phi(\mathbf{s})\right)$ as CP $i$'s throughput.

{\lemma For any $s'_i > s_i$, let $\mathbf{s'} = (s'_i, \mathbf{s}_{-i})$ and $\mathbf{s} = (s_i, \mathbf{s}_{-i})$ for some fixed strategy profile $\mathbf{s}_{-i}$. For any price $p$,
$\phi(\mathbf{s'}) \geq \phi(\mathbf{s}), \ \theta_i(\mathbf{s'}) \geq \theta_i(\mathbf{s}) $ and $\theta_j(\mathbf{s'}) \leq \theta_j(\mathbf{s}), \ \forall j\neq i$.\label{lemma:subsidy_effect}}

Lemma \ref{lemma:subsidy_effect} states that if a CP $i$ unilaterally increases its subsidy $s_i$, its throughput $\theta_i$ and the system utilization $\phi$ increase; however, the throughput of any other CP decreases.
It shows that the subsidization mechanism creates a competitive game among the CPs, where each CP $i$ could use subsidy $s_i$ to maximize its utility $U_i$.
Notice that besides the policy constraint $q$, each CP's optimal subsidy strategy depends on all other CPs' strategies as well as the ISP's price $p$.
In this section, we study the policy implications of subsidization and therefore, we often assume a fixed price $p$ for the access ISP.
This setting corresponds to a competitive access market where the ISPs cannot easily manipulate prices, or a case where the ISP's price is regulated. We will study an ISP's pricing strategy, its revenue and the policy implication on the system welfare in the next section.

{\definition[Nash equilibrium] Given any fixed ISP price $p$ and policy $q$, a strategy profile $\mathbf{s}$ is a {\em Nash equilibrium } if each $s_i$ solves CP $i$'s utility maximization problem:
\begin{eqnarray*}
\max~~  & U_i(s_i; \mathbf{s}_{-i}) = (v_i-s_i)\theta_i(\mathbf{s}) \\
\mathrm{subject~to~}~  & 0\leq s_i\leq q.
\end{eqnarray*}\label{definition:Nash_equilibrium}}
We characterize the Nash equilibrium of the subsidization competition game in the following theorem.
{\theorem[Characterization] For any fixed  $p$ and  $q$, a strategy profile $\mathbf s$ is a Nash equilibrium only if
\[  \quad s_i = \min\{\tau_i(\mathbf s),q\}, \quad \forall \ i\in\cal N,\]
where each $\tau_i$ denotes a threshold for CP $i$, defined by
\begin{equation}
\displaystyle \tau_i(\mathbf s) \triangleq 
\left(v_i-s_i\right)\epsilon_{s_i}^{m_i}\left(1+\epsilon_{\phi}^{\lambda_i}\epsilon_{m_i}^{\phi}\right).\label{equation:tau}
\end{equation}
In particular, $v_i \leq \left({\partial \theta_i}/{\partial s_i}\right)^{-1} \theta_i$ holds if $s_i=0$.
Moreover, if $U_i(\mathbf{s})=U_i(s_i, {\mathbf s}_{-i})$ is concave in $s_i$ for all $i\in\cal N$, then the above conditions are also sufficient.
\label{theorem:Nash_equilibrium}}

Theorem \ref{theorem:Nash_equilibrium} characterizes the Nash equilibrium in two ways. First, it indicates that
any CP $i$ does not subsidize if its profitability $v_i$ is lower than $(\partial \theta_i/\partial s_i)^{-1}\theta_i$. Second, if the policy constraint is not tight, i.e., $s_i<q$, the equilibrium subsidy $s_i$ can be characterized by its profit margin $v_i-s_i$ and the elasticity metrics $\epsilon_{s_i}^{m_i}$, $\epsilon_{\phi}^{\lambda_i}$ and $\epsilon_{m_i}^{\phi}$. Equation (\ref{equation:tau}) can also be written as $\tau_i(\mathbf{s}) = (v_i-s_i) \epsilon_{s_i}^{\theta_i} = (v_i-s_i) \epsilon_{s_i}^{m_i}\left(1+\epsilon_{m_i}^{\lambda_i}\right)$.
This characterization also implies that a CP would subsidize its users more if 1) its traffic profitability $v_i$ (and therefore its profit margin $v_i-s_i$) increases, or 2) its throughput elasticity $\epsilon_{s_i}^{\theta_i}$ or user demand elasticity of subsidy $\epsilon_{s_i}^{m_i}$ increases.

By Assumption \ref{assumption:utilization_throughput} and \ref{assumption:user_base}, each $U_i(\mathbf{s})$ is differentiable in $\mathbf{s}$. For any strategy profile $\mathbf{s}$, we define $\mathbf{u}(\mathbf{s}) = \{u_i(\mathbf{s}): i\in\cal N\}$ as the vector of marginal utilities, where each $u_i(\mathbf{s})$ denotes the marginal utility of CP $i$, defined as $u_i(\mathbf{s}) \triangleq \partial U_i(\mathbf{s})/\partial s_i$.
Next, we characterize the uniqueness of Nash equilibrium based on a condition on the marginal utilities $\mathbf{u}(\mathbf{s})$.

{\theorem[Uniqueness] For any fixed price $p$ and policy $q$, if for any distinct pair of feasible strategy profiles $\mathbf{s' \neq s}$, there always exist a CP $i\in\cal N$ such that
\begin{equation} (s'_i - s_i)\left(u_i(\mathbf{s'})-u_i(\mathbf{s})\right)<0,
\label{equation:stricly_monotone}\end{equation} then there always exists a unique Nash equilibrium. \label{theorem:unique_Nash_equilibrium}}

Technically, the above sufficient condition for uniqueness requires $-\mathbf{u}$ to be a {\em $P$-function} \cite{More73}.
Notice that this uniqueness property holds for any subset ${\cal Q}\subset [0,q]^{|\cal N|}$ of the strategy space if condition (\ref{equation:stricly_monotone}) holds in the domain ${\cal Q} $.


\subsection{Dynamics of Subsidies in Equilibrium}
Since a Nash equilibrium is defined under a fixed price $p$ and policy $q$ by Definition (\ref{definition:Nash_equilibrium}), we use $\mathbf{s}=\mathbf{s}(p,q)$ to indicate a Nash equilibrium under $p$ and $q$.
Equation (\ref{equation:tau}) in Theorem \ref{theorem:Nash_equilibrium} hinted that a CP with higher profitability subsidizes users more in an equilibrium.
We formally show this as follows. 

{\theorem[Profitability Effect on Subsidy] If a CP $i$'s profitability increases from $v_i$ to $\hat v_i$ unilaterally, and under the condition (\ref{equation:stricly_monotone}) of Theorem \ref{theorem:unique_Nash_equilibrium}, $\mathbf{s}$ and $\mathbf{\hat s}$ are the corresponding Nash equilibria, then we must have $\hat s_i \geq s_i$. \label{theorem:subsidy_valuation_monotonicity}}

If a CP's profitability increases, Theorem \ref{theorem:subsidy_valuation_monotonicity} tells that it will increase its subsidy in equilibrium, and hints that its throughput will increase by Lemma \ref{lemma:subsidy_effect}. 
Through subsidization, CPs can respond to their profitability and influence throughput.

After analyzing the impact of CPs' profitability, we study how an equilibrium $\mathbf{s}(p,q)$ is influenced by the price $p$ and policy $q$.
We define $\cal N_{-}$ and $\cal N_{+}$ as the set of CPs whose subsidies are $0$ and $q$, respectively, and $\widetilde{\cal  N} = {\cal N}\backslash (\cal N_{-}\cup \cal N_{+})$ as the remaining set of CPs whose subsidies are strictly positive and less than $q$. We define $\mathbf{\tilde s}$ and $\mathbf{\tilde u}$ as the subsidies and marginal utilities of the set $\widetilde{\cal  N}$ of CPs.

{\theorem[Equilibrium Dynamics] If a Nash equilibrium $\mathbf{s}$ satisfies 1) $i\in\widetilde{\cal  N}$ if $u_i(\mathbf{s})=0$, and 2) the condition (\ref{equation:stricly_monotone}) for a neighborhood of $\mathbb{R}^{|\cal N|}$ centered at $\mathbf{s}$ in the strategy space, then in a neighborhood of $(p,q)$, a unique Nash equilibrium is a differentiable function $\mathbf{s}(p,q)$, satisfying
\begin{equation}
\frac{\partial s_i}{\partial q} =  \left\{
\begin{array}{cl}
\displaystyle 0 & \text{if } \ i\in{\cal N}_-; \\
\displaystyle 1 & \text{if } \ i\in{\cal N}_+; \\
\displaystyle  -\sum_{k\in\widetilde{\cal N}} \psi_{ik} \sum_{j\in{\cal N}_+} \frac{\partial u_k}{\partial s_j}, & \text{if } \ i\in\widetilde{\cal N},
\end{array} \right. \label{equation:partial_q}
\end{equation}
\begin{equation}
\quad \text{and} \quad \frac{\partial s_i}{\partial p} =  \left\{
\begin{array}{cl}
 0 & \text{if } \ i\notin\widetilde{\cal N}; \\
 \displaystyle -\sum_{k\in\widetilde{\cal N}} \psi_{ik} \frac{\partial u_k}{\partial p}, & \text{if } \ i\in\widetilde{\cal N},
\end{array} \right. \label{equation:partial_p}
\end{equation}
where $\Psi = \{\psi_{ij}\} \triangleq (\nabla_{\mathbf{\tilde s}} \mathbf{\tilde u})^{-1}$, i.e., $\psi_{ij}$ is at the $i$th row and $j$th column of the inverse of the Jacobian matrix $\nabla_{\mathbf{\tilde s}} \mathbf{\tilde u}$.
\label{theorem:equilibrium_dynamics}}

Theorem \ref{theorem:equilibrium_dynamics} states that a marginal change in the ISP's price or the regulatory policy will not affect the behavior of the set ${\cal N}_-$ and ${\cal N}_+$ of CPs: the CPs who do not subsidize remain the same and the CPs who subsidize the amount $q$ keep subsidizing at the maximum level as $q$ increases.
When $p$ changes, the set $\widetilde{\cal N}$ of CPs readjust subsidies among themselves in a new equilibrium such that their marginal utilities remain zero in (\ref{equation:partial_p}); when $q$ changes, since the set ${\cal N}_+$ of CPs increase subsidies accordingly, their impacts have to be counted in the new equilibrium in (\ref{equation:partial_q}).
By applying the KKT condition to Definition \ref{definition:Nash_equilibrium}, we know $u_i(\mathbf{s})=0$ for all $i\in\widetilde{\cal  N}$, and the first condition of Theorem \ref{theorem:equilibrium_dynamics} is a bit stronger which guarantees that the equilibrium $\mathbf{s}$ is {\em regular}, i.e., locally differentiable. The second condition only assumes (\ref{equation:stricly_monotone}) locally, which guarantees the local uniqueness of equilibrium in a neighborhood centered at $\mathbf{s}$ by Theorem \ref{theorem:unique_Nash_equilibrium}.

{\corollary[Deregulation] Under a fixed ISP price $p$ and the conditions of Theorem \ref{theorem:equilibrium_dynamics}, in a neighborhood of $q$, we can express $\phi(q)$ and $R(q)$ as the system utilization and the ISP's revenue under the Nash equilibrium $\mathbf{s}(q)$. If $\mathbf u$ is off-diagonally monotone, i.e., $\partial u_i(\mathbf{s})/\partial s_j\geq 0$ for all $i\neq j$, 
\[  \frac{\partial \phi}{\partial q}\geq 0, \quad  \frac{\partial R}{\partial q}\geq 0 \quad \text{and} \quad \frac{\partial s_i}{\partial q} \geq 0, \quad \forall i\in\cal N.\]
\label{corollary:utilization_monotonicity}}

The off-diagonally monotone property makes $-\mathbf{u}$ to be a Leontief type \cite{Gale65} of $P$-matrix, which guarantees the stability of a macroeconomic system due to Wassily Leontief \cite{Hawkins49}.
Intuitively, it assumes although the utility of a CP $i$ decreases when other CPs increase their subsidies, its marginal benefit of subsidizing its own users, i.e., $u_i$, increases.
Corollary \ref{corollary:utilization_monotonicity} states that under this stability condition, when allowed to subsidize more, CPs will increase subsidies, which will lead to an increase in the system utilization and the ISP's revenue.

{ \it \bf Regulatory Implications:}  Under a competitive or price regulated access market, deregulation of subsidization encourages CPs to subsidize users (by Theorem \ref{theorem:subsidy_valuation_monotonicity} and \ref{theorem:equilibrium_dynamics}), and consequently increases the system utilization and the ISPs' revenue (by Corollary \ref{corollary:utilization_monotonicity}).
CPs' user populations $\mathbf{m}$ increase because the usage charges become cheaper after subsidization (by Assumption \ref{assumption:user_base}), but the induced higher utilization (by Theorem \ref{theorem:capacity_user_base_effect}) might cause congestion.
The total system throughput also increases (implied by Corollary \ref{corollary:utilization_monotonicity} as $\phi$ increases), while the throughput of certain CPs might decrease.
However, this is caused by the negative network externality, i.e., the physics of congestion of shared capacity.
Although it might be unfavorable for the CPs whose traffic is more congestion sensitive in the short term,
as the ISPs improve profit margins from higher utilization, they will have more incentives to expand capacities so as to accommodate more traffic and relieve congestion in the long term.

\section{ISP Revenue and System Welfare}\label{sec:revenue}

Although the policy $q$, the price $p$ and the subsidies $\mathbf{s}$ all impact the system,  these decisions are not made simultaneously and independently.
The system is fundamentally driven by a regulatory policy $q$, under which the ISP determines its price $p(q)$ and then the CPs respond with strategies $\mathbf{s}(p,q)$. Theorem \ref{theorem:equilibrium_dynamics} characterizes the dynamics of $\mathbf{s}(p,q)$ and Corollary \ref{corollary:utilization_monotonicity} characterizes the impact of policy $q$ when $p$ is fixed. In practice, ISPs often use price to optimize revenue, and by Theorem \ref{theorem:equilibrium_dynamics} the subsidies $\mathbf{s}$ are also influenced by $p$.
In this section, we study the impact of the ISP's price on its revenue and the impact of policy on the system welfare when the response of ISP's pricing is taken into account.

\subsection{Impact of ISP's Pricing on Its Revenue}
Under a fixed policy $q$, by Theorem \ref{theorem:equilibrium_dynamics}, we can write $\mathbf{s}(p)$ as the induced subsidies under price $p$. Thus, the induced system utilization is $\phi(\mathbf{s}(p))$ and the ISP's induced revenue is $R(p) = p\sum_{i\in\cal N}\theta_i(p) = p\sum_{i\in\cal N}m_i(p-s_i(p))\lambda_i(\phi(\mathbf{s}(p)))$.
{\theorem[Marginal Revenue] If the ISP's revenue is differentiable at $p$, then its marginal revenue is
\begin{equation}
\frac{dR(p)}{dp} = \sum_{i\in\cal N} \theta_i + \Upsilon\sum_{i\in\cal N}  \epsilon_p^{m_i} \theta_i, \label{equation:marginal_R}
\end{equation}
where $\Upsilon$ and the $m_i$-elasticity of price $p$ satisfy
\[ \Upsilon = 1+\sum_{j\in\cal N} \epsilon_{m_j}^{\lambda_j} \quad \text{and} \quad \epsilon_p^{m_i} = \frac{p}{m_i}\frac{dm_i}{dt_i}\left(1-\frac{\partial s_i}{\partial p}\right).\]
\label{theorem:opt_price}}

Theorem \ref{theorem:opt_price} characterizes the change in revenue as the price varies and isolates the effect of subsidization into the elasticities $\epsilon_{p}^{m_i}$, where ${\partial s_i}/{\partial p}$ plays a role. It also generalizes the case of one-sided pricing, i.e., ${\partial s_i}/{\partial p}=0$ for all $i\in\cal N$. The second term in (\ref{equation:marginal_R}) measures the price's impact on the aggregate throughput, i.e., $p({\partial \theta}/{\partial p})$, which can be factorized by $\Upsilon$, a parameter determined by the physical model in Figure \ref{figure:network_model}, because  $\epsilon_{m_j}^{\lambda_j}$ can be decomposed as $\epsilon_{m_j}^{\phi}\epsilon_{\phi}^{\lambda_j}$ and
\begin{equation}
\epsilon_{m_j}^{\phi}\epsilon_{\phi}^{\lambda_j}= \frac{m_j}{\lambda_j} \frac{d\lambda_j}{d\phi}\frac{\partial\phi}{\partial m_j} = m_j\frac{d\lambda_j}{d\phi}\left(\frac{dg}{d\phi}\right)^{-1}.
\label{equation:epsilon}\end{equation}

\subsection{Impact of Regulatory Policy on Welfare}
By Theorem \ref{theorem:opt_price}, we understand how the ISP's pricing affects its revenue. We assume that the ISP would adapt its price $p$ as a function of a given policy $q$. The next theorem captures the impact of the policy $q$ on the system states, i.e., $\phi$, $m_i$ and $\theta_i$, where the responses of both the ISP's price $p(q)$ and the CPs' subsidies $\mathbf{s}(p,q)$ are taken into account.

{\theorem[Policy Effect] If the ISP's price is a differentiable function $p(q)$ of policy $q$ and $\mathbf{s}$ is a Nash equilibrium of $p(q)$ and $q$, under the conditions of Theorem \ref{theorem:equilibrium_dynamics}, in a neighborhood of $q$, we can express the Nash equilibrium as $\mathbf{s}(q) \triangleq \mathbf{s}\left(p(q),q\right)$.
The corresponding system utilization $\phi(q)$, user population $m_i(q)$ and throughput $\lambda_i(q)$ satisfy
\begin{equation}
\frac{dm_i}{dq} = \frac{dm_i}{dt_i} \frac{dt_i}{dq} = \frac{dm_i}{dt_i} \left(\left(1-\frac{\partial s_i}{\partial p}\right)\frac{dp}{dq} - \frac{\partial s_i}{\partial q}\right), \label{equation:dmi_dq}
\end{equation}
\begin{equation}
\frac{d\phi}{dq} = \left(\frac{dg}{d\phi}\right)^{-1} \sum_{i\in\cal N} \frac{dm_i}{dq}\lambda_i \quad \text{and} \quad \frac{d\lambda_i}{dq} = \frac{d\lambda_i}{d\phi} \frac{d\phi}{dq}. \label{equation:dphi_dq}
\end{equation}
Any CP $i$'s throughput $\theta_i$ increases with $q$ if and only if
\begin{equation} {\epsilon_{t_i}^{m_i}\epsilon_q^{t_i}}/{\epsilon_\phi^{\lambda_i}} < - \epsilon_q^\phi. \label{inequality:q}\end{equation}
\label{theorem:regulation_effect}}
Theorem \ref{theorem:regulation_effect} shows that the policy effect on utilization $d\phi/dq$ in (\ref{equation:dphi_dq}) and the condition (\ref{inequality:q}) for throughput $\theta_i$ have similar forms as the price effect $d\phi/dp$ in (\ref{equation:phi_p}) and the condition (\ref{inequality:price_effect}). 
This is because the policy effect is carried out by its impact on the user populations, i.e., $\epsilon_{q}^{m_i}=\epsilon_{t_i}^{m_i}\epsilon_q^{t_i}$, via its impact on the price and subsidies. By Assumption \ref{assumption:utilization_throughput} and \ref{assumption:user_base}, both ${\epsilon_\phi^{\lambda_i}}$ and $\epsilon_{t_i}^{m_i}$ are negative. 
Inequality (\ref{inequality:q}) tells that a CP's throughput decreases if any only if $(-\epsilon_{t_i}^{m_i})(-\epsilon_q^{t_i})<-{\epsilon_\phi^{\lambda_i}}\epsilon_q^\phi$, which implies that the subsidy could help increase throughput via $(-\epsilon_q^{t_i})$; however, $(-\epsilon_{t_i}^{m_i})$ could be decreased due to the ISP's increasing price, which will reduce the CP's throughput.


{\corollary[Policy Impact on Welfare] Under the conditions of Theorem \ref{theorem:regulation_effect}, let $W(q) \triangleq \sum_{i\in\cal N} \theta_i(q)v_i$ define the system welfare. Suppose ${d \phi}/{d q}$ in (\ref{equation:dphi_dq}) is positive, then the marginal welfare ${d W}/{d q}$ is positive if and only if
\[ \sum_{i\in\cal N} \frac{w_i}{\sum_{k\in\cal N} w_k} v_i > \sum_{i\in\cal N} -\epsilon_{m_i}^{\lambda_i} v_i, \quad \text{where} \quad w_i \triangleq \lambda_i \frac{d m_i}{dq}.\]
\label{corollary:welfare}}

We measure the system welfare in terms of the gross profit $W=\sum_{i\in\cal N}{\theta_i v_i}$ of all the CPs for two reasons.
First, it internalizes the subsidy transfer from CPs to ISP. Second, because CP profits are often positively correlated to their values to users, it also serves an estimate for user welfare.
In Corollary \ref{corollary:welfare}, $w_iv_i$ can be interpreted as the increase in welfare due to the policy's impact on the user population $m_i$, and therefore, the left side of the inequality represents the normalized increase in welfare due to the changes in the user populations. The right side of the inequality represents the normalized decrease in welfare due to the policy's impact on each average throughput $\lambda_i$ via $m_i$. Corollary \ref{corollary:welfare} states that the welfare increases if and only if the increasing component is larger than the decreasing component. Notice that the decreasing component only depends on the physical characteristics (\ref{equation:epsilon}) of the CPs, while the weight $w_i/\sum_{j\in\cal N}w_j$ for each $v_i$ tends to be large if $v_i$ is large, because profitable CPs will have stronger tendencies to subsidize their users so as to attract a larger population $m_i$.

To understand the policy effect more intuitively, we perform a numerical evaluation as follows. We use the same setting $m_i(t_i)=e^{-\alpha_it_i}$, $\lambda_i(\phi)=e^{-\beta_i\phi}$ and $\Phi(\theta,\mu)=\theta/\mu$ as in Section \ref{sec:model}.
We model $8$ types of CPs with $\alpha_i,\beta_i \in \{2,5\}$ and $v_i\in\{0.5,1\}$. By Lemma \ref{lemma:CP_characteristics}, each CP represents the aggregation of a group of CPs with similar characteristics of traffic and user demand. In each of the following figures, we vary the policy $q$ at $5$ levels from $0$ to $2.0$ and vary the ISP's price $p$ from $0$ to $2$ along the x-axis. When $q=0$, the figures show the baseline case where subsidization is not allowed.

\begin{figure}[!h]
\centering
\includegraphics[width=.5\columnwidth]{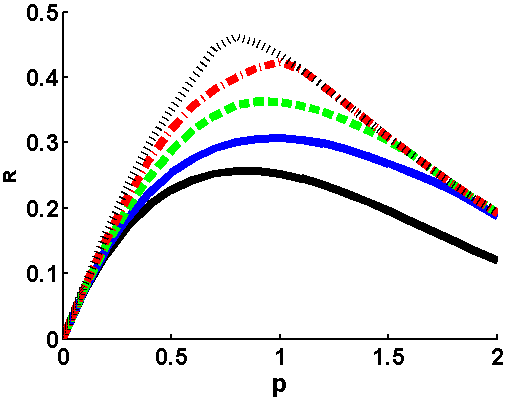}\includegraphics[width=.5\columnwidth]{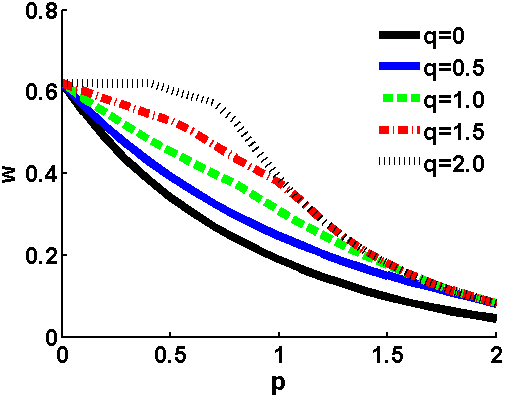}
\caption{ISP's Revenue $R$ and System Welfare $W$.}
\label{figure:Revenue}
\end{figure}

Figure \ref{figure:Revenue}  plots the ISP's revenue $R$ (left) and the system welfare $W$ (right). We observe that under any fixed price $p$, both the ISP's revenue and the system welfare are higher when the CPs' subsidization is less regulated, i.e., $q$ is large.
However, if the deregulation of subsidization will trigger a higher ISP price, it also shows that the system welfare will decrease with the price $p$ under any fixed policy $q$.

\begin{figure*}[!ht]
\centering
\includegraphics[width=.99\linewidth]{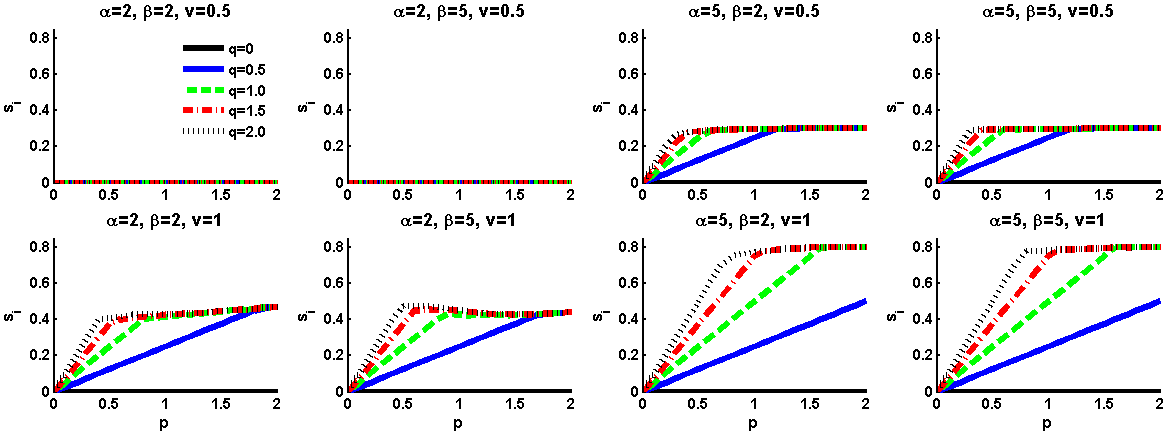}
\caption{The subsidies $s_i$ of the $8$ types of CPs under equilibrium.}
\label{figure:s08}
\end{figure*}

\begin{figure*}[!ht]
\centering
\includegraphics[width=.99\linewidth]{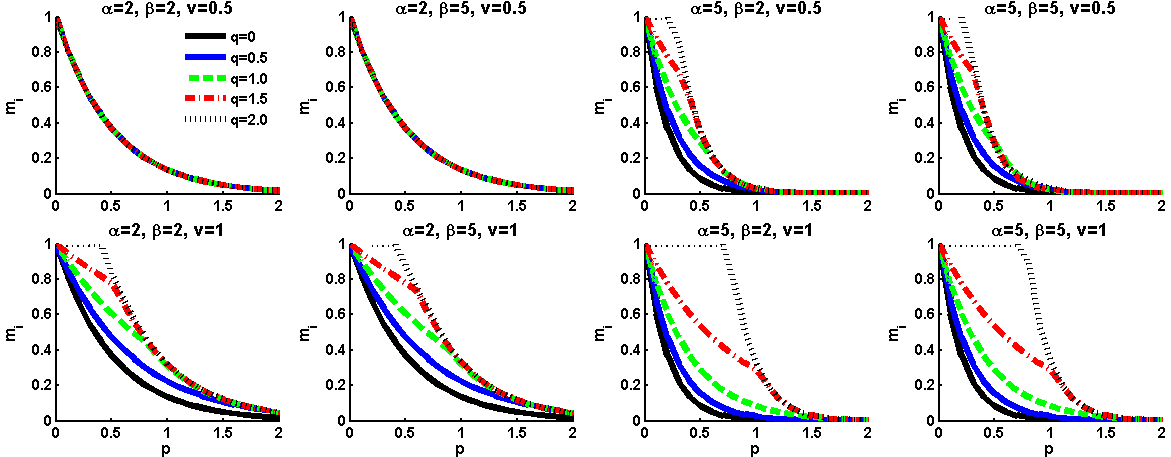}
\caption{The user population $m_i$ of the $8$ types of CPs under equilibrium.}
\label{figure:m08}
\end{figure*}

\begin{figure*}[!ht]
\centering
\includegraphics[width=.99\linewidth]{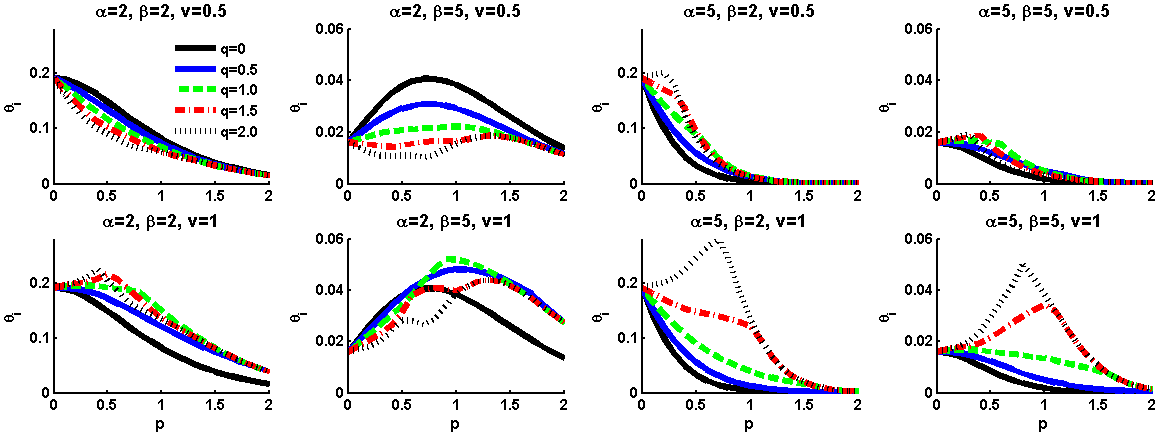}
\caption{The throughput $\theta_i$ of the $8$ types of CPs under equilibrium.}
\label{figure:theta08}
\end{figure*}

\begin{figure*}[!ht]
\centering
\includegraphics[width=.99\linewidth]{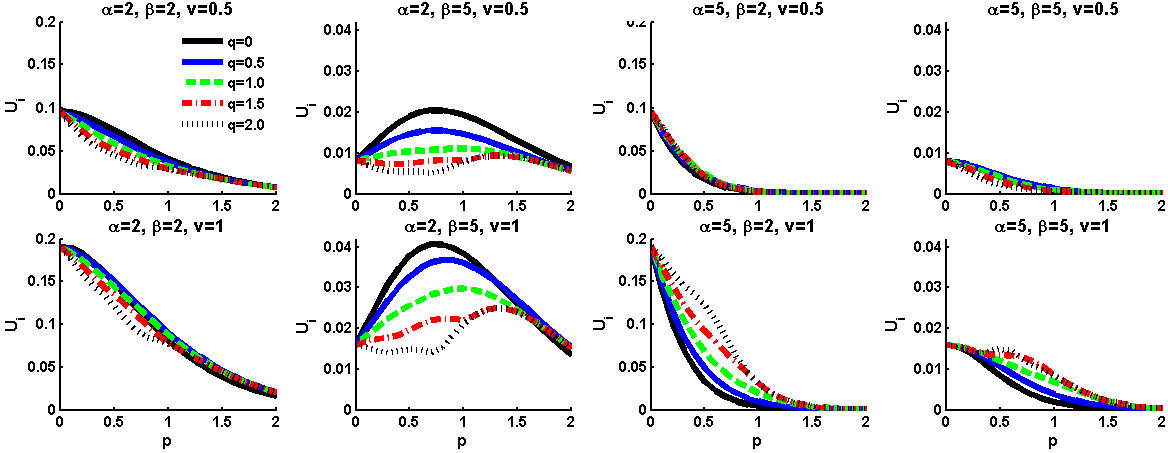}
\caption{The utility $U_i$ of the $8$ types of CPs under equilibrium.}
\label{figure:U08}
\end{figure*}

Figure \ref{figure:s08} plots the subsidy $s_i$ of the $8$ types of CPs under equilibrium.
We observe that the CPs that have a higher profitability, i.e., $v_i=1$ in the lower $4$ sub-figures, or a higher demand elasticity, i.e., $\alpha_i=5$ in the right $4$ sub-figures, provide much higher subsidies compared to their counterparts. We also observe that when the price $p$ is small, except for the two CPs with $\alpha_i=2$ and $v_i=0.5$, most CPs want to subsidize at the maximum level $q$ constrained by the policy; however, when $p$ increases, subsidies may stay flat and then decrease due to the decrease in profit margin. By comparing Figure \ref{figure:s08} with Figure \ref{figure:Revenue}, we observe that when $q=2$, the ISP maximizes its revenue by setting $p$ a bit less than $1$, where the CPs' subsidies are kept at a high level.

Figure \ref{figure:m08} plots the corresponding user population $m_i$ of the $8$ types of CPs under equilibrium.
We observe that when $p$ increases, the user populations of the CPs with higher demand elasticity in the right $4$ sub-figures decrease steeper than those with lower demand elasticity. By comparing the low-value CPs in the upper sub-figures with the high-value CPs in the lower sub-figures,
we observe that the user populations of the high-value CPs decrease much slower so that they can retain higher user populations via higher subsidies.
In all cases, we observe that CPs always obtain higher user population under a more relaxed policy $q$.

Figure \ref{figure:theta08} plots the individual throughput $\theta_i$ of the $8$ types of CPs under equilibrium. In Figure \ref{figure:theta08}, we observe that the CPs with higher profitability, i.e., $v_i=1$, or lower congestion elasticity, i.e., $\beta_i=2$, achieve higher throughput compared to their counterparts.
When comparing the throughput with that of the baseline, i.e., $q=0$, we observe that the
CPs with high profitability achieve higher throughput, with the only exception for the case $(\alpha_i, \beta_i, v_i)=(2,5,1)$ when $p$ is small.
As $p$ is small, the system attracts large user demand and is relatively congested. Because this CP is more congestion sensitive, i.e., $\beta_i$ is large, and by Corollary \ref{corollary:utilization_monotonicity}, subsidization will further increase the system utilization, its throughput will be reduced.
By comparing Figure \ref{figure:theta08} with Figure \ref{figure:Revenue}, we observe that with higher throughput for the CPs of higher profitability, the system welfare increases when the policy $q$ is relaxed.

Figure \ref{figure:U08} plots the resulting utility $U_i$ of the $8$ types of CPs under equilibrium.
As the individual CP's utility is defined as $U_i=(v_i-s_i)\theta_i$, each $U_i$ shows similar trends as $\theta_i$ in Figure \ref{figure:theta08}, which is scaled by its profit margin $v_i-s_i$. When $q$ increases, we observe that 1) CPs with high demand elasticity and value, i.e., $\alpha_i=5$ and $v_i=1$, achieve higher utility via higher subsidies, user population and throughput, and 2) CPs with low demand elasticity and high congestion elasticity, i.e., $\alpha_i=2$ and $\beta_i=5$, achieve lower utility due to the lower achievable throughput. The utilities of other CPs under different policies are comparable.


{ \it \bf Regulatory Implications:}  In a deregulated market where CPs are allowed to subsidize their users, highly profitable CPs will provide greater subsidies (by Theorem \ref{theorem:subsidy_valuation_monotonicity}) and obtain more users (by Assumption \ref{assumption:user_base}), achieving higher throughput (by Theorem \ref{theorem:capacity_user_base_effect}) and leading to a higher system welfare (by Corollary \ref{corollary:welfare}).
Due to negative network externality, some CPs may obtain lower throughput; however, it can be resolved when ISPs obtain investment incentives and expand capacity.
Nonetheless, deregulation of subsidization might trigger an increase in the ISP's price, leading to a decrease in the system welfare (shown in Figure \ref{figure:Revenue}).
High ISP prices also discourage user demand (by Assumption \ref{assumption:user_base}) 
and consequently decrease their throughput (shown in Figure \ref{figure:theta08}).
Our results suggest that the subsidization competition will increase the ISP's revenue and system welfare (by Corollary \ref{corollary:utilization_monotonicity}); however, price regulations might be needed if the access market is not competitive and the ISP's price is set too high.

\section{Issues and Model Limitations}\label{sec:issues}

Subsidization works with the usage-based pricing schemes of the access ISPs.
For mobile providers, implementing data caps and accounting for data traffic consumptions are common in practice.
Broadband providers in countries like India and Australia have been using usage-based pricing, and major U.S. ISPs like Verizon \cite{Segall} and AT\&T \cite{Taylor} have also adopted usage-based pricing after gaining support from the FCC \cite{schatz10fcc,Johnson13fcc}.
To differentiate content and enable subsidization from different CPs, platforms such as  FreeBand \cite{FreeBand} have been developed to provide data traffic statistics to mobile carriers so as to lower the bills of consumers, based on the type of applications.
AT\&T's recent sponsored data \cite{ATT14} plan allows CPs to fully subsidize their users.
Although our model allows partial subsidization, its technical feasibility is the same as the proposed sponsored data applications.


Deregulation of subsidization might reduce the throughput of certain CPs, e.g., the upper-left ones in Figure \ref{figure:theta08}.
These CPs either do not have incentives to subsidize, because their users are not price-sensitive, i.e., the elasticity $\epsilon_{s_i}^{m_i}$ is small, or they cannot afford to subsidize, because their profitability $v_i$ is low. In the former case, the decrease in throughput is mainly due to a higher utilization of the system, which will be relieved in the long term when the ISPs expand capacities.
Startup companies with low profits are of the latter case. The decrease in throughput is mainly due to the ISP's high price that limits the user demand, and therefore, regulators might want to regulate ISP's monopolistic pricing or introduce competition, e.g., municipal infrastructures \cite{ma13public}.
However, if they are promising, we also expect that venture capital would provide the funding source for them to subsidize their users and achieve their potential high profitability.
After all, we believe in a transparent and competitive market where users can choose CPs based on quality and prices, and the businesses drive the evolution of the Internet ecosystem.

Although our analyses imply that a deregulation of subsidization will provide CPs more freedom to subsidize and increase the competitiveness of the content market, regulators might still want to impose policies that prevent discriminatory subsidization of the access ISPs. For example, Comcast's Xbox XFinity \cite{ComcastXFINITY} service enables its users to access on-demand content on their Xbox devices, which does not count against their $250$GB data cap. By implicitly subsidizing their users, this service favors the ISP's vertically integrated business over other content competitors, e.g., Netflix.
Some CPs work with access ISPs to provide better services and cheaper prices for their users, e.g., Google Apps for Verizon \cite{GoogleVerizon}; however, this might also inversely affect the competitors of those favored CPs.
We believe that the subsidization option should be identical, and should be given to all CPs equally. Through this, the access ISPs would treat CPs more neutrally and the subsidization would create a transparent and uniform platform for the CPs to compete.

Our analytical framework is based on a macroscopic equilibrium model that captures the interactions among an access ISP, a set of CPs and end-users via the CPs' subsidization and the ISP's capacity and pricing decisions.
Our theoretical results are mostly qualitative, which provides understanding and predictions of the changes in the system states when some of the driving factors change. The limitation is that it might not be able to capture short-term off-equilibrium types of system dynamics, where players' decisions are not rational or optimal.
Our numerical evaluations are limited to a styled model of CPs to capture qualitative trends; however, more detailed experiments and validations could be challenging, because market data are needed so as to obtain the characteristics of the CPs, e.g., their profitability and elasticities, and an ISP needs to execute such a plan in the market. With the emerging sponsored data plan from AT\&T, we expect this type of market data could be available for regulatory authorities as well as research communities.
This study focuses on a single access ISP; however, we believe that competition between ISPs will also incentivize them to adopt subsidization schemes, through which users can obtain subsidized services.
Finally, our result shows that under the subsidization mechanism, the access ISP could increase utilization and revenue, which will further provide investment incentives for them to expand capacity. We do not have sufficient space to study the ISP's capacity planning decision in detail, which is a direction of our future work.

\section{Conclusions}
The Internet is facing a dilemma of increasing traffic demand and decreasing incentives for the access ISPs to expand capacity, because service prioritization and two-sided pricing are not allowed under the net neutrality principles.
We propose to allow CPs to voluntarily subsidize the data usage costs incurred at the access ISPs for their end-users.
Our solution modularizes the issue along the tussle boundary such that the physical network is still kept neutral while the innovative competition among CPs is enabled at higher service layers.
Through our macroscopic modeling and analyses, we show that a deregulated subsidization policy will incentivize highly profitable CPs to subsidize users and increase the system welfare and the revenue of the access ISPs.
With the improved profit margins, the access ISPs would obtain more investment incentives under subsidization. However, the deregulation of subsidization might trigger an ISP to increase price, which would decrease the system welfare as well as the throughput of startup CPs. We suggest that price regulation might be needed if the access market is not competitive enough and the price is too high.
Because subsidization creates a feedback channel for CPs to return value to the access ISPs and realign the created value and investment in the value chain, we believe that subsidization competition could vitalize neutral Internet for the future.

\appendix
\section{Proofs}
{\bf Proof of Lemma \ref{lemma:unique_utilization}:}
By Assumption \ref{assumption:utilization_throughput}, we know $\Theta(\phi,\mu)$ is strictly increasing in $\phi$ and $\lambda_i(\phi)$ is strictly decreasing in $\phi$; therefore, $g(\phi)=\Theta(\phi,\mu)-\sum_{i\in\cal N} m_i \lambda_i(\phi)$ is strictly increasing in $\phi$.
Because $\lim_{\phi\rightarrow 0} \Theta(\phi,\mu) = 0$ and $\lambda_i(0)> 0$ for all $i\in\cal N$, $lim_{\phi\rightarrow 0}g(\phi) < 0$; because $\lim_{\phi\rightarrow \infty}\lambda_i(\phi) = 0$ for all $i\in\cal N$, $lim_{\phi\rightarrow \infty}g(\phi) =\lim_{\phi\rightarrow \infty} \Theta(\phi,\mu) > 0$. As a result, $g(\phi)=0$ must have a unique solution.

Finally, when $g(\phi)=0$, $\Theta(\phi,\mu)=\sum_{i\in\cal N} m_i \lambda_i(\phi)$, which is equivalent to Equation (\ref{equation:utilization}) of Definition \ref{definition:utilization}.

\done

{\bf Proof of Lemma \ref{lemma:CP_characteristics}:} Let $\kappa = m_j / m_i$. Given CP $i$'s characterization $\lambda_i(\cdot)$, we verify that $\lambda_j(\phi)=\kappa^{-1}\lambda_i(\phi)$ for all $\phi>0$. This is because when $\lambda_j(0)=\kappa^{-1}\lambda_i(0)$, the above implies $\epsilon_{\phi}^{\lambda_j}(\phi) = \left(\frac{d\lambda_j}{d\phi}\right)\left(\frac{\phi}{\lambda_j}\right) = \left(\frac{1}{\kappa}\frac{d\lambda_i}{d\phi}\right)\left(\frac{\phi}{\kappa^{-1}\lambda_i}\right) = \epsilon_{\phi}^{\lambda_i}(\phi)$ for all $\phi>0$. Thus, $\theta_j(\phi) = m_j \lambda_j(\phi) = \kappa m_i \kappa^{-1}\lambda_i(\phi) = m_i\lambda_i(\phi) = \theta_i(\phi)$ for all $\phi>0$.
Finally, because given $\phi$, the throughput of CP $i$ and $j$ are the same, the system will induce the same utilization $\phi$, and therefore, $\phi$ is also the unique system utilization when CP $i$ is replaced by CP $j$.

\done

{\bf Proof of Theorem \ref{theorem:capacity_user_base_effect}:}
Because $g(\phi)$ is strictly increasing in $\phi$ by Lemma \ref{lemma:unique_utilization} and by Equation (\ref{equation:dg}), we have
\[ \frac{dg}{d\phi} = \frac{\partial \Theta}{\partial \phi} - \sum_{k\in\cal N}m_k \frac{d\lambda_k}{d\phi}>0. \]
When considering the capacity effect, we write the system utilization $\phi(\mu)$ as a function of the system capacity. By Lemma \ref{lemma:unique_utilization}, $g(\phi(\mu)) = \Theta(\phi(\mu),\mu) - \sum_{k\in\cal N}m_k\lambda_k(\phi(\mu)) = 0$.
By taking derivative of $\mu$ on both sides, we obtain
\[ \frac{\partial \Theta}{\partial \phi} \frac{\partial \phi}{\partial \mu} +\frac{\partial \Theta}{\partial \mu} - \sum_{k\in\cal N}m_k \frac{d\lambda_k}{d\phi} \frac{\partial \phi}{\partial \mu} = 0, \]
which is equivalent to 
\[ \left(\frac{\partial \Theta}{\partial \phi} - \sum_{k\in\cal N}m_k \frac{d\lambda_k}{d\phi}\right) \frac{\partial \phi}{\partial \mu} = -\frac{\partial \Theta}{\partial \mu} \quad \text{or} \quad  \frac{dg}{d\phi}\frac{\partial \phi}{\partial \mu} = -\frac{\partial \Theta}{\partial \mu}. \]
Since $\Theta(\phi,\mu)$ is increasing in $\mu$, $\partial \Theta/\partial \mu>0$, which implies
\[ \frac{\partial \phi}{\partial \mu} = -\left(\frac{dg}{d\phi}\right)^{-1}\frac{\partial \Theta}{\partial \mu}<0. \]
When considering the user effect, we write the system utilization $\phi({\bf m})$ as a function of the user populations. By Lemma \ref{lemma:unique_utilization}, $g(\phi({\bf m})) = \Theta(\phi({\bf m}),\mu) - \sum_{k\in\cal N}m_k\lambda_k(\phi({\bf m})) = 0$.
By taking derivative of $m_i$ on both sides, we obtain
\[ \frac{\partial \Theta}{\partial \phi} \frac{\partial \phi}{\partial m_i}  - \sum_{k\in\cal N}m_k \frac{d\lambda_k}{d\phi} \frac{\partial \phi}{\partial m_i} - \lambda_i = 0, \]
which is equivalent to
\[ \left(\frac{\partial \Theta}{\partial \phi} - \sum_{k\in\cal N}m_k \frac{d\lambda_k}{d\phi}\right) \frac{\partial \phi}{\partial m_i} = \lambda_i  \quad \text{or} \quad  \frac{dg}{d\phi}\frac{\partial \phi}{\partial m_i} = \lambda_i. \]
Since $\frac{dg}{d\phi}>0$, the above implies $\frac{\partial \phi}{\partial m_i} = \left(\frac{dg}{d\phi}\right)^{-1}\lambda_i>0$.
Finally, the results for $\partial \theta_i/\partial \mu$, $\partial \theta_i/\partial m_i$ and $\partial \theta_j/\partial m_i$ can be derived by using $\theta_i(\phi) = m_i \lambda_i(\phi)$ and chain rules, and applying the above results for $\partial \phi/\partial \mu$ and $\partial \phi/\partial m_i$.

\done

{\bf Proof of Theorem \ref{theorem:price_effect}:}
When considering the price effect, we write the system utilization $\phi(p)=\phi({\bf m}(p))$ as a function of the price. By Lemma \ref{lemma:unique_utilization}, $g(\phi(p)) = \Theta(\phi(p),\mu) - \sum_{k\in\cal N}m_k(p)\lambda_k(\phi(p)) = 0$.
By taking derivative of $p$ on both sides, we obtain
\[ \frac{\partial \Theta}{\partial \phi} \frac{\partial \phi}{\partial p}  - \sum_{k\in\cal N} \left( \frac{dm_k}{dp}\lambda_k + m_k \frac{d\lambda_k}{d\phi} \frac{\partial \phi}{\partial p} \right) = 0, \]
which is equivalent to 
\[ \left(\frac{\partial \Theta}{\partial \phi} - \sum_{k\in\cal N}m_k \frac{d\lambda_k}{d\phi}\right) \frac{\partial \phi}{\partial p} = \sum_{k\in\cal N}  \frac{dm_k}{dp}\lambda_k. \]
By Assumption \ref{assumption:user_base}, $m_k(p)$ is decreasing in $p$, which implies
\[\frac{\partial \phi}{\partial p} = \left(\frac{dg}{d\phi}\right)^{-1} \sum_{k\in\cal N}  \frac{dm_k}{dp}\lambda_k \leq 0. \]
By taking derivatives of $p$ on $\theta_i(p) = m_i(p)\lambda_i(\phi(p))$,
\[ \frac{d\theta_i}{dp} > 0 \Leftrightarrow {\frac{dm_i}{dp}}{{\lambda_i}} + m_i \frac{d\lambda_i}{d\phi}\frac{\partial \phi}{\partial p}>0, \]
which is equivalent to
\[ \frac{1}{m_i}{\frac{dm_i}{dp}} > -\frac{1}{{\lambda_i}}\frac{d\lambda_i}{d\phi}\frac{\partial \phi}{\partial p} \quad \text{or} \quad {\epsilon_p^{m_i}}/{\epsilon_\phi^{\lambda_i}} < - \epsilon_p^\phi.\]
Finally,
\[ \frac{d\theta}{dp} = \sum_{i\in\cal N} \frac{d\theta_i}{dp} = \sum_{i\in\cal N} {\frac{dm_i}{dp}}{{\lambda_i}} + m_i \frac{d\lambda_i}{d\phi}\frac{\partial \phi}{\partial p}. \]
By substituting $\partial\phi/\partial p$ of (\ref{equation:phi_p}) into the above, we obtain
\[ \frac{d\theta}{dp} = \sum_{i\in\cal N} {\frac{dm_i}{dp}}{{\lambda_i}} \left(1+ \left(\frac{dg}{d\phi}\right)^{-1} \sum_{k\in\cal N} {\frac{d\lambda_k}{d\phi}}{{m_k}} \right).\]
By substituting $dg/d\phi$ of (\ref{equation:dg}) into the above, we obtain (\ref{equation:dtheta_dp}).

\done

{\bf Proof of Lemma \ref{lemma:subsidy_effect}:}
Because $s'_i>s_i$, $t'_i = p-s'_i < p-s_i =t_i$ and by Assumption \ref{assumption:user_base}, $m'_i = m_i(t'_i) \geq m_i(t_i) = m_i$.

By Theorem \ref{theorem:capacity_user_base_effect}, we know that $\partial \phi/\partial m_i >0$, and therefore, $\phi(\mathbf{s}')\geq \phi(\mathbf{s})$. Also by Theorem \ref{theorem:capacity_user_base_effect}, we know that $\partial \theta_i/\partial m_i>0$ and $\partial \theta_j/\partial m_i<0$ for all $j\neq i$. As a result, $\theta_i(\mathbf{s}')\geq \theta_i(\mathbf{s})$ and $\theta_j(\mathbf{s}')\leq \theta_j(\mathbf{s})$ for all $j\neq i$, which implies $U_j(\mathbf{s}') = v_j \theta_j(\mathbf{s}') \leq v_j\theta_j(\mathbf{s}) = U_j(\mathbf{s})$ for all $j\neq i$.

\done

{\bf Proof of Theorem \ref{theorem:Nash_equilibrium}:} If $\mathbf{s}$ is a Nash equilibrium, each $s_i \in [0,q]$ maximizes
$U_i(s_i; \mathbf{s}_{-i}) = (v_i-s_i)m_i(p-s_i)\lambda_i\left(\phi\left(\mathbf{s}\right)\right)$.
By Karush-Kuhn-Tucker condition, we can derive that
\begin{equation}
\frac{\partial U_i}{\partial s_i}  \left\{
\begin{array}{cl}
\leq 0 & \text{if } \ s_i=0; \\
\geq 0 & \text{if } \ s_i=q; \\
= 0 & \text{if } \ 0<s_i<q;
\end{array} \right. \quad \forall \ i\in\cal N.
\label{equation:KKT}
\end{equation}
For CP $i$ with $s_i=0$, the above condition tells that
\[ \frac{\partial U_i}{\partial s_i} = (v_i-s_i)\frac{\partial \theta_i}{\partial s_i} - \theta_i = v_i\frac{\partial \theta_i}{\partial s_i} - \theta_i \leq 0.\]
By Lemma \ref{lemma:subsidy_effect}, we know ${\partial \theta_i}/{\partial s_i}>0$ and the above implies
$v_i \leq \left({\partial \theta_i}/{\partial s_i}\right)^{-1} \theta_i$.
By taking partial derivatives of $s_i$ on $U_i=(v_i-s_i)m_i\lambda_i$, we deduce that
\[ \frac{\partial U_i}{\partial s_i} = (v_i-s_i)\left(m_i\frac{\partial \lambda_i}{\partial s_i}+\lambda_i\frac{\partial m_i}{\partial s_i}\right)  - m_i\lambda_i. \]
Dividing $m_i\lambda_i$ on both sides of (\ref{equation:KKT}), we have
\begin{equation*}
(v_i-s_i)\left(\frac{1}{\lambda_i}\frac{\partial \lambda_i}{\partial s_i}+\frac{1}{m_i}\frac{\partial m_i}{\partial s_i}\right)  \left\{
\begin{array}{cl}
\leq 1 & \text{if } \ s_i=0; \\
\geq 1 & \text{if } \ s_i=q; \\
= 1 & \text{if } \ 0<s_i<q;
\end{array} \right.
\end{equation*}
When we multiply $s_i$ on both sides of the above, the left hand side becomes
\[(v_i-s_i)\left(\epsilon_{s_i}^{m_i}+\epsilon_{s_i}^{\lambda_i}\right) =(v_i-s_i)\left(\epsilon_{s_i}^{m_i}+\epsilon_{\phi}^{\lambda_i}\epsilon_{m_i}^{\phi}\epsilon_{s_i}^{m_i}\right) = \tau_i(\bf s).\]
When $s_i=0$, $\tau_i({\bf s}) = 0 =s_i$, and therefore, we have
\begin{equation*}
 \tau_i({\bf s})  \left\{
\begin{array}{cl}
\geq s_i & \text{if } \ s_i=q; \\
= s_i & \text{if } \ 0 \leq s_i<q,
\end{array} \right.  \quad {\text or} \quad s_i = \min\{\tau_i(\mathbf s),q\}.
\end{equation*}
Finally, if $U_i(\mathbf{s})$ is concave in $s_i$ for all $i\in\cal N$, local optimality also guarantees the global optimality and therefore, the above first order necessary conditions become sufficient condition for $\mathbf{s}$ being a Nash equilibrium.

\done

{\bf Proof of Theorem \ref{theorem:unique_Nash_equilibrium}:}
The condition (\ref{equation:stricly_monotone}) implies that $U_i(s_i; \mathbf{s}_{-i})$ is concave in $s_i$ for any $\mathbf{s}_{-i}$. Because the strategy space $[0,q]$ is compact, it guarantees the existence of Nash equilibrium.
Suppose there exists two distinct Nash equilibria $\mathbf{\hat s}$ and $\mathbf{\tilde s}$. By concavity of $U_i(s_i; \mathbf{s}_{-i})$ in $s_i$ and the maximum principle,  for any $i\in\cal N$ and any $x_i\in[0,q]$,
\[ (x_i - \hat s_i)u_i(\mathbf{\hat s}) \leq 0, \quad \text{and} \quad (x_i - \tilde s_i)u_i(\mathbf{\tilde s}) \leq 0.\]
By substituting $\mathbf{x} = \mathbf{\tilde s}$ in the first inequality and $\mathbf{x} = \mathbf{\hat s}$ in the second inequality, we deduce for any $i\in\cal N$,
\[ (\tilde s_i - \hat s_i)u_i(\mathbf{\hat s}) \leq 0, \quad \text{and} \quad (\hat s_i - \tilde s_i)u_i(\mathbf{\tilde s}) \leq 0.\]
By adding the above inequalities, we further deduce that
\[ (\tilde s_i - \hat s_i)(u_i(\mathbf{\hat s}) - u_i(\mathbf{\tilde s})) \leq 0, \quad \forall i\in\cal N.\]
The above is equivalent to $(\tilde s_i - \hat s_i)(u_i(\mathbf{\tilde s}) - u_i(\mathbf{\hat s})) \geq 0$ for all $i\in\cal N$, which contradicts the condition (\ref{equation:stricly_monotone}).

\done

{\bf Proof of Theorem \ref{theorem:subsidy_valuation_monotonicity}:}
For any $j\in{\cal N}\backslash \{i\}$, we show that
\[\left(\hat s_j - s_j\right)\left(u_j(\mathbf{\hat s})-u_j(\mathbf{s})\right)\geq 0. \]
Because the above is symmetric in $\mathbf{s}$ and $\mathbf{\hat s}$, without loss of generality, we can consider the cases of $s_j\leq \hat s_j$. By using the KKT condition of (\ref{equation:KKT}), we have the following four cases.
\begin{enumerate}
\item $s_j = \hat s_j$: $\left(\hat s_j - s_j\right)\left(u_j(\mathbf{\hat s})-u_j(\mathbf{s})\right)=0$.
\item $s_j=0$ and $\hat s_j \in (0,q)$: $u_j(\mathbf{s})<0$ and $u_j(\mathbf{\hat s})=0$; and therefore, the above becomes $\hat s_j \left(-u_j(\mathbf{s})\right)\geq 0$.
\item $s_j=0$ and $\hat s_j = q$: $u_j(\mathbf{s})<0$ and $u_j(\mathbf{\hat s})>0$; and therefore, the above becomes $q\left(u_j(\mathbf{\hat s})-u_j(\mathbf{s})\right)\geq 0$.
\item $s_j \in (0,q)$ and $\hat s_j = q$: $u_j(\mathbf{s})=0$ and $u_j(\mathbf{\hat s})>0$; and therefore, the above becomes $\hat (q - s_j) u_j(\mathbf{\hat s})\geq 0$.
\end{enumerate}
By the condition (\ref{equation:stricly_monotone}), we deduce $\left(\hat s_i - s_i\right)\left(u_i(\mathbf{\hat s})-u_i(\mathbf{s})\right) < 0$.
Suppose $\hat s_i < s_i$, the above implies that $u_i(\mathbf{\hat s}) > u_i(\mathbf{s})$.
However, if $\hat s_i < s_i$, we must have $s_i>0$ and by the KKT condition of (\ref{equation:KKT}), we have
$u_i(\mathbf{s}) \geq 0$, which further implies that $u_i(\mathbf{\hat s}) > u_i(\mathbf{s})\geq 0$.
Again by the KKT condition of (\ref{equation:KKT}), $u_i(\mathbf{\hat s}) > 0$ implies $\hat s = q$ which contradicts with the assumption $\hat s_i < s_i$. Consequently, we must have $\hat s_i \geq s_i$.
\done

{\bf Proof of Theorem \ref{theorem:equilibrium_dynamics}:}
The condition (\ref{equation:stricly_monotone}) of Theorem \ref{theorem:unique_Nash_equilibrium} implies the local concavity of the utility functions. By Proposition 1.4.2 of \cite{book-VI}, the Nash equilibrium $\mathbf{s}$ can be equivalently characterized as the solution of a variational inequality, denoted as $VI(F,K)$, where $F\triangleq -\mathbf{u}$ and $K\triangleq [0,q]^{|\cal N|}$. $\mathbf{s}\in\cal K$ is a solution of $VI(F,K)$ if $(\mathbf{x}-\mathbf{s})^TF(\mathbf{s})\geq 0$ for all $\mathbf{x}\in K$. We apply the sensitivity analysis \cite{Tobin86, Dafermos88} of variational inequalities to obtain the dynamics of the Nash equilibrium $\mathbf{s}(p,q)$ as the price $p$ or the policy $q$ changes.

For each CP $i$, the constraint $s_i\in[0,q]$ can be written as two linear constraints $g^-_i(\mathbf{s})  \triangleq  s_i \geq 0$ and $g^+_i(\mathbf{s}) \triangleq  q - s_i \geq 0$.
For any $\mathbf{s}$, the set of binding constraints are
\begin{equation}
{\cal G} = \{g^-_i(\mathbf{s}):i\in{\cal N_-}\} \cup \{g^+_i(\mathbf{s}):i\in{\cal N_+}\}.
\label{equation:constaint_set}
\end{equation}
Because ${\cal N_-}\cap{\cal N_+}=\emptyset$, the gradients of the binding constraints are linearly independent.
Because $u_i(\mathbf{s}) = 0$ implies $s_i \in \widetilde{\cal N}$ for all $i\in\cal N$, by the KKT condition of (\ref{equation:KKT}), we know that the strict complementary slackness condition of Theorem 3.1 of \cite{Tobin86} holds.
By Theorem \ref{theorem:unique_Nash_equilibrium}, the Nash equilibrium is locally unique, and therefore, by Theorem 3.1 of \cite{Tobin86}, in a neighborhood of $(p,q)$, the Nash equilibrium can be written as a differentiable function $\mathbf{s}(p,q)$.

Let $m \triangleq |\cal N_-|+|\cal N_+|$ be the number of binding constraints. We define $G$ as a $m\times n$ matrix, whose $i$th row is the gradient of the $i$th binding constraint with respect to $\mathbf{s}$.
We define $E_p$ and $E_q$ as $m\times 1$ vectors, whose $i$th component are the partial derivatives of the $i$th binding constraint with respect to $p$ and $q$, respectively.
Following the framework of \cite{Dafermos88}, we define $Q = G^T(GG^T)^{-1}G$ and $M$ to be an $n\times m$ matrix satisfying $MG=Q$ and $QM=M$.
Given the set $\cal G$ of linear constraints in (\ref{equation:constaint_set}), we can deduce that $Q = \{q_{ij}\}$ is a diagonal matrix satisfying $q_{ii}=1$ if $i\notin \widetilde{\cal N}$ and $q_{ii}=0$ otherwise, and $M=G^T$.

When $q$ is the sensitivity parameter, $E_q = \{e_j\}$ satisfies that $e_j=1$ if the $j$th constraint is a type of $g^+_i(\mathbf{s}) = q-s_i$ constraint and $e_j=0$ otherwise.
Because the constraints $\cal G$ are linear, Theorem 3.1 of \cite{Dafermos88} implies that $Q \nabla_q \mathbf{s}(q) = -ME_q$ and $(I-Q)(\nabla_{\mathbf{s}} F \nabla_{q} \mathbf{s}(q) + \nabla_{q} F) = \mathbf{0}$.
Because when multiplying $\nabla_q \mathbf{s}(q)$ by $Q$, the values associated with the non-binding strategies vanish, the first equation implies
\begin{equation}
\partial s_i / \partial q = 0, \ \ \forall i\in{\cal N}_- \ \ \text{and} \ \ \partial s_i / \partial q = 1, \ \ \forall i\in{\cal N}_+.
\label{equation:tmp1}
\end{equation}
Because $F\triangleq -\mathbf{u}$ and $\mathbf{u}$ does not depend on $q$, the second equation can be written as
$(I-Q)\nabla_{\mathbf{s}} \mathbf{u} \nabla_{q} \mathbf{s} = \mathbf{0}$. Because when multiplied by $I-Q$, the components associated with the binding strategies vanish, by substituting (\ref{equation:tmp1}) into the above, we obtain 
$\nabla_{\mathbf{\tilde s}} \mathbf{\tilde u} \nabla_q \mathbf{\tilde s} + \nabla_{\mathbf{s}} \mathbf{\tilde u} E_q = \mathbf{0}$.
By the condition (\ref{equation:stricly_monotone}) of Theorem \ref{theorem:unique_Nash_equilibrium}, we deduce that $F\triangleq -\mathbf{u}$ is a {\em $P$-function} \cite{More73}. If we restrict to the CPs in $\widetilde{\cal N}$, we know that the Jacobian
$\nabla_{\mathbf{\tilde s}} (\mathbf{-\tilde u})$ is a {\em $P$-matrix} \cite{More73}, which is always non-singular. Therefore, we deduce $\nabla_q \mathbf{\tilde s} = - (\nabla_{\mathbf{\tilde s}} \mathbf{\tilde u})^{-1}\nabla_{\mathbf{s}} \mathbf{\tilde u} E_q = - \Psi\nabla_{\mathbf{s}} \mathbf{\tilde u} E_q$, the same as the third case of (\ref{equation:partial_q}).

When taking $p$ as the sensitivity parameter and applying Theorem 3.1 of \cite{Dafermos88}, we can obtain $Q \nabla_p \mathbf{s}(p) = -ME_p$ and $(I-Q)(\nabla_{\mathbf{s}} F \nabla_{p} \mathbf{s}(p) + \nabla_{p} F) = \mathbf{0}$.
Because the constraints in $\cal G$ do not depend on $p$, $E_p$ is a zero vector, therefore, the first equation implies that $\partial s_i/\partial p = 0$ for all $i\notin \widetilde{\cal N}$. By substituting $F \triangleq -\mathbf{u}$ into the second equation, we have
$(I-Q)(\nabla_{\mathbf{s}} \mathbf{u} \nabla_{p} \mathbf{s} + \nabla_{p} \mathbf{u}) = \mathbf{0}$.
Similarly, by substituting $\partial s_i/\partial p = 0$ for all $i\notin \widetilde{\cal N}$ into the above, we obtain $\nabla_{\mathbf{\tilde s}} \mathbf{\tilde u} \nabla_p \mathbf{\tilde s} + \nabla_{p} \mathbf{\tilde u} = \mathbf{0}$. Because $\nabla_{\mathbf{\tilde s}} (\mathbf{-\tilde u})$ is non-singular, it is equivalent to the second case of (\ref{equation:partial_p}) in a matrix form as
\[\nabla_p \mathbf{\tilde s} = - (\nabla_{\mathbf{\tilde s}} \mathbf{\tilde u})^{-1}\nabla_{p} \mathbf{\tilde u}= - \Psi \nabla_{p} \mathbf{\tilde u}.\]
\done

{\bf Proof of Corollary \ref{corollary:utilization_monotonicity}:}
By (\ref{equation:partial_q}) of Theorem \ref{theorem:equilibrium_dynamics}, $\partial s_i/\partial q \geq 0$ is immediate for $i\notin\widetilde{\cal N}$. For the CPs in $\widetilde{\cal N}$, by Theorem \ref{theorem:equilibrium_dynamics}, we have $ \nabla_q \mathbf{\tilde s} = - (\nabla_{\mathbf{\tilde s}} \mathbf{\tilde u})^{-1}\nabla_{\mathbf{s}} \mathbf{\tilde u} E_q = - \Psi\nabla_{\mathbf{s}} \mathbf{\tilde u} E_q$.
Because $\mathbf{u}$ is off-diagonally monotone, we know that $\nabla_{\mathbf{\tilde s}} (-\mathbf{\tilde u})$ is a $P$-matrix with non-positive off-diagonal entries or an $M$-matrix \cite{Poole74}, which implies that all the entries of $-\Psi$ are nonnegative.
By Theorem \ref{theorem:equilibrium_dynamics}, we know that $\nabla_{\mathbf{s}} \mathbf{\tilde u} E_q$ is a vector of $\{\partial u_i/\partial s_j : i\in {\widetilde{\cal N}}, j\in {\cal N}_+\}$. Because $\mathbf{u}$ is off-diagonally monotone, the components of $\nabla_{\mathbf{s}} \mathbf{\tilde u} E_q$ are all non-negative, and therefore, $\nabla_q \mathbf{\tilde s} = -\Psi \nabla_{\mathbf{s}} \mathbf{\tilde u} E_q \geq \mathbf{0}$.


Now we have shown that $\partial s_i/\partial q \geq 0$ for all $i\in{\cal N}$. Since all the subsidies are monotonic, by Lemma \ref{lemma:subsidy_effect}, we deduce that $\partial \phi/\partial q \geq 0$. Finally, because the ISP's revenue can be written as $R = p\Theta(\phi,\mu)$ and $\partial \Theta(\phi,\mu)/\partial \phi \geq 0$,  we have
\[ \frac{\partial R}{\partial q} = \frac{\partial R}{\partial \phi} \frac{\partial \phi}{\partial q} = p\frac{\partial \Theta(\phi,\mu)}{\partial \phi} \frac{\partial \phi}{\partial q} \geq 0. \]
\done

{\bf Proof of Theorem \ref{theorem:opt_price}:} By Equation (\ref{equation:phi_mi}) and (\ref{equation:phi_p}), we deduce
\[ \frac{\partial \phi}{\partial p} =  \frac{\partial \phi}{\partial m_i} \frac{1}{\lambda_i} \sum_{j\in\cal N} \frac{\partial m_j}{\partial p} \lambda_j, \quad \forall i\in\cal N.\]
By substituting the above into $\displaystyle \frac{\partial \theta_i}{\partial p} = \frac{\partial m_i}{\partial p}\lambda_i + m_i\frac{d \lambda_i}{d p} \frac{\partial \phi }{\partial p}$,
we obtain $\displaystyle \frac{\partial \theta_i}{\partial p} = \frac{\partial m_i}{\partial p}\lambda_i + m_i\frac{d \lambda_i}{d p} \frac{\partial \phi}{\partial m_i} \frac{1}{\lambda_i} \sum_{j\in\cal N} \frac{\partial m_j}{\partial p} \lambda_j$, or

\begin{equation*}
\begin{aligned}
\epsilon_p^{\theta_i} \theta_i = & p\frac{\partial \theta_i}{\partial p} = \frac{p}{m_i}\frac{\partial m_i}{\partial p}\theta_i + m_i\frac{d \lambda_i}{d p} \frac{\partial \phi}{\partial m_i} \frac{1}{\lambda_i} \sum_{j\in\cal N} \frac{p}{m_j}\frac{\partial m_j}{\partial p} \theta_j \\
    = &  \epsilon_p^{m_i} \theta_i + \epsilon_{m_i}^{\phi} \epsilon_{\phi}^{\lambda_i} \sum_{j\in\cal N} \epsilon_p^{m_j} \theta_j
\end{aligned}
\end{equation*}
Because $R(p) = p\sum_{i\in\cal N}\theta_i(p)$, the marginal revenue is
\begin{equation*}
\begin{aligned}
\frac{dR}{dp} = & \sum_{i\in\cal N} \theta_i + p \sum_{i\in\cal N} \frac{\partial \theta_i}{\partial p} =  \sum_{i\in\cal N} \theta_i + \sum_{i\in\cal N}\epsilon_p^{\theta_i}  \theta_i \\
= & \sum_{i\in\cal N} \theta_i + \sum_{i\in\cal N} \epsilon_p^{m_i} \theta_i + \epsilon_{m_i}^{\phi} \epsilon_{\phi}^{\lambda_i} \sum_{j\in\cal N} \epsilon_p^{m_j} \theta_j \\
= & \sum_{i\in\cal N} \theta_i + \sum_{i\in\cal N} \epsilon_p^{m_i} \theta_i +   \sum_{i\in\cal N} \epsilon_{m_i}^{\phi} \epsilon_{\phi}^{\lambda_i} \sum_{j\in\cal N} \epsilon_p^{m_j} \theta_j \\
= & \sum_{i\in\cal N} \theta_i + \Upsilon\sum_{i\in\cal N}  \epsilon_p^{m_i} \theta_i.
\end{aligned}
\end{equation*}
\vspace{-0.2in}

\done

{\bf Proof of Theorem \ref{theorem:regulation_effect}:} By Theorem \ref{theorem:equilibrium_dynamics}, the Nash equilibrium is locally unique, i.e., $s_i(q)\triangleq s_i\left(p(q),q\right)$, we deduce
\begin{equation}\frac{ds_i}{dq} = \frac{\partial  s_i}{\partial q} + \frac{\partial s_i}{\partial p} \frac{dp}{dq}. \label{equation:dsi_dq}\end{equation}
Thus, we can obtain (\ref{equation:dmi_dq}) by substituting the above into
\[\frac{dm_i}{dq} = \frac{dm_i}{dt_i} \frac{dt_i}{dq} = \frac{dm_i}{dt_i} \left(\frac{dp}{dq} - \frac{ds_i}{dq} \right).\]
Similarly, we can obtain (\ref{equation:dphi_dq}) by substituting (\ref{equation:phi_mi}) into
\[ \frac{\partial \phi}{\partial q} = \sum_{i\in\cal N} \frac{\partial \phi}{\partial m_i} \frac{d m_i}{dq}.\]
Finally, $\displaystyle \frac{\partial \theta_i}{\partial q}>0$ can be written as $\displaystyle {\frac{dm_i}{dq}}{{\lambda_i}} + m_i \frac{d\lambda_i}{d\phi}\frac{d \phi}{d q}>0$.
By Assumption \ref{assumption:utilization_throughput}, $\epsilon_{\phi}^{\lambda_i}<0$ and because
$\epsilon_q^{\theta_i} = \epsilon_q^{m_i} + \epsilon_q^{\lambda_i} = \epsilon_q^{t_i}\epsilon_{t_i}^{m_i} + \epsilon_q^{\phi}\epsilon_{\phi}^{\lambda_i}$ and ${\partial \theta_i}/{\partial q}>0$ is equivalent to $\epsilon_q^{\theta_i}>0$, \[ \epsilon_q^{\theta_i} > 0 \Leftrightarrow   \epsilon_q^{t_i}\epsilon_{t_i}^{m_i} > -\epsilon_q^{\phi}\epsilon_{\phi}^{\lambda_i} \Leftrightarrow   \epsilon_q^{t_i}\epsilon_{t_i}^{m_i}/\epsilon_{\phi}^{\lambda_i} < -\epsilon_q^{\phi}.\]
\done

{\bf Proof of Corollary \ref{corollary:welfare}:} Because $W(q) \triangleq \sum_{i\in\cal N} \theta_i(q)v_i$,
\[\frac{dW}{dq} = \sum_{i\in\cal N}  m_i \frac{d\lambda_i}{dq} v_i + \sum_{i\in\cal N} \lambda_i \frac{dm_i}{dq} v_i.\]
By substituting $w_i \triangleq \lambda_i \frac{dm_i}{dq}$ and (\ref{equation:dphi_dq}) into the above,
\begin{equation*}
\begin{aligned}
\frac{dW}{dq} = & \sum_{i\in\cal N}  m_i \frac{d\lambda_i}{d\phi} \frac{d\phi}{dq} v_i + \sum_{i\in\cal N} w_i v_i \\
= & \sum_{i\in\cal N}  m_i \frac{d\lambda_i}{d\phi} \left(\frac{dg}{d\phi}\right)^{-1} \left(\sum_{j\in\cal N} w_j \right) v_i + \sum_{i\in\cal N} w_i v_i.
\end{aligned}
\end{equation*}
Because $d\phi/dq>0$, by (\ref{equation:dphi_dq}) we know that $\sum_{j\in\cal N} w_j>0$. Therefore, $dW/dq>0$ is equivalent to
\[\sum_{i\in\cal N}  m_i \frac{d\lambda_i}{d\phi} \left(\frac{dg}{d\phi}\right)^{-1}  v_i + \sum_{i\in\cal N} \frac{w_i}{\sum_{j\in\cal N} w_j } v_i > 0.\]
By substituting $\left({dg}/{d\phi}\right)^{-1} = \lambda_i^{-1} \frac{\partial \phi}{\partial m_i}$ into the above, 
\[\sum_{i\in\cal N}  \frac{m_i}{\lambda_i} \frac{d\lambda_i}{d\phi} \frac{\partial \phi}{\partial m_i}  v_i + \sum_{i\in\cal N} \frac{w_i}{\sum_{j\in\cal N} w_j } v_i > 0,\]
which is equivalent to $\sum_{i\in\cal N} \frac{w_i}{\sum_{k\in\cal N} w_k} v_i > \sum_{i\in\cal N} -\epsilon_{m_i}^{\lambda_i} v_i$.

\done 

\end{document}